\documentclass[  reprint,
  amsmath,amssymb,
  aps,
  prd,
  citenumbers
]{revtex4-1}

\usepackage{graphicx}
\usepackage{dcolumn}

\usepackage[utf8]{inputenc}
\usepackage[T1]{fontenc}
\usepackage{etoolbox}
\usepackage{amsmath}
\usepackage{amssymb}
\usepackage{bbm}
\usepackage[colorlinks=false, linkcolor=black]{hyperref}
\usepackage{braket}
\usepackage{siunitx}
\usepackage{placeins}
\usepackage{wrapfig}
\usepackage[justification=justified,singlelinecheck=false]{caption}
\usepackage{enumitem} 
\usepackage{newtxtext,newtxmath}
\usepackage{ragged2e}
\usepackage{hyperref}
\usepackage{float}

\setcitestyle{etalmode=truncate, maxnames=2, minnames=1}

\makeatletter
\patchcmd{\@makecaption}
  {\fnum@figure~\fname@caption}
  {\fnum@figure~\fname@caption\par}
  {}{}

\makeatother

\newcommand{\bs}[1]{\boldsymbol{\mathsf{#1}}}
\newcommand{\mc}[1]{\mathcal{#1}}

\makeatletter
\def\@email#1#2{%
 \endgroup
 \patchcmd{\titleblock@produce}
  {\frontmatter@RRAPformat}
  {\frontmatter@RRAPformat{\produce@RRAP{*#1\href{mailto:#2}{#2}}}\frontmatter@RRAPformat}
  {}{}
}%
\makeatother

\begin{document}

\preprint{AIP/123-QED}

\title[Toolchain]{Toolchain for shuttling trapped-ion qubits in segmented traps}
\author{A. Conta}
 \email{conta@uni-mainz.de}
\author{S. Bogino}%
\author{F. Köhncke}%
\author{F. Schmidt-Kaler}%
\author{U. G. Poschinger}
\affiliation{%
QUANTUM, Institut für Physik, Staudingerweg 7, 55128 Mainz
}%

\date{\today}

\begin{abstract}
Scalable trapped-ion quantum computing requires fast and reliable transport of ions through complex, segmented radiofrequency trap architectures without inducing excessive motional excitation. We present a numerical toolchain for the systematic generation of time-dependent electrode voltages enabling fast, low-excitation ion shuttling in segmented radiofrequency traps. Based on a model of the trap electrode geometry, the framework combines an electrostatic field solver, efficient unconstrained optimization, waveform postprocessing, and dynamical simulations of ion motion to compute voltage waveforms that realize prescribed transport trajectories while respecting experimental constraints such as voltage limits and bandwidth. The toolchain supports arbitrary trap geometries, including junctions and multi-zone layouts, and allows for the flexible incorporation of optimization objectives. We provide a detailed assessment of the accuracy of the framework by investigating its numerical stability and by comparing measured and predicted secular frequencies. The framework is optimized for numerical performance, enabling rapid numerical prototyping of trap architectures of increasing complexity. As application examples, we apply the framework to the transport of a potential well along a linear, uniformly segmented trap, and we compute a solution for shuttling a potential well around the corner of an X-type trap junction. The presented approach provides an extensible and highly efficient numerical foundation for designing and validating transport protocols in current and next-generation trapped-ion processors.
\end{abstract}

\maketitle

\makeatletter
\renewcommand\l@subsubsection[2]{} 
\makeatother
\tableofcontents

\section{Introduction}
\begin{figure}[!ht]
    \includegraphics[width=\columnwidth,trim={0.5cm 5cm 1.5cm 4cm},clip]{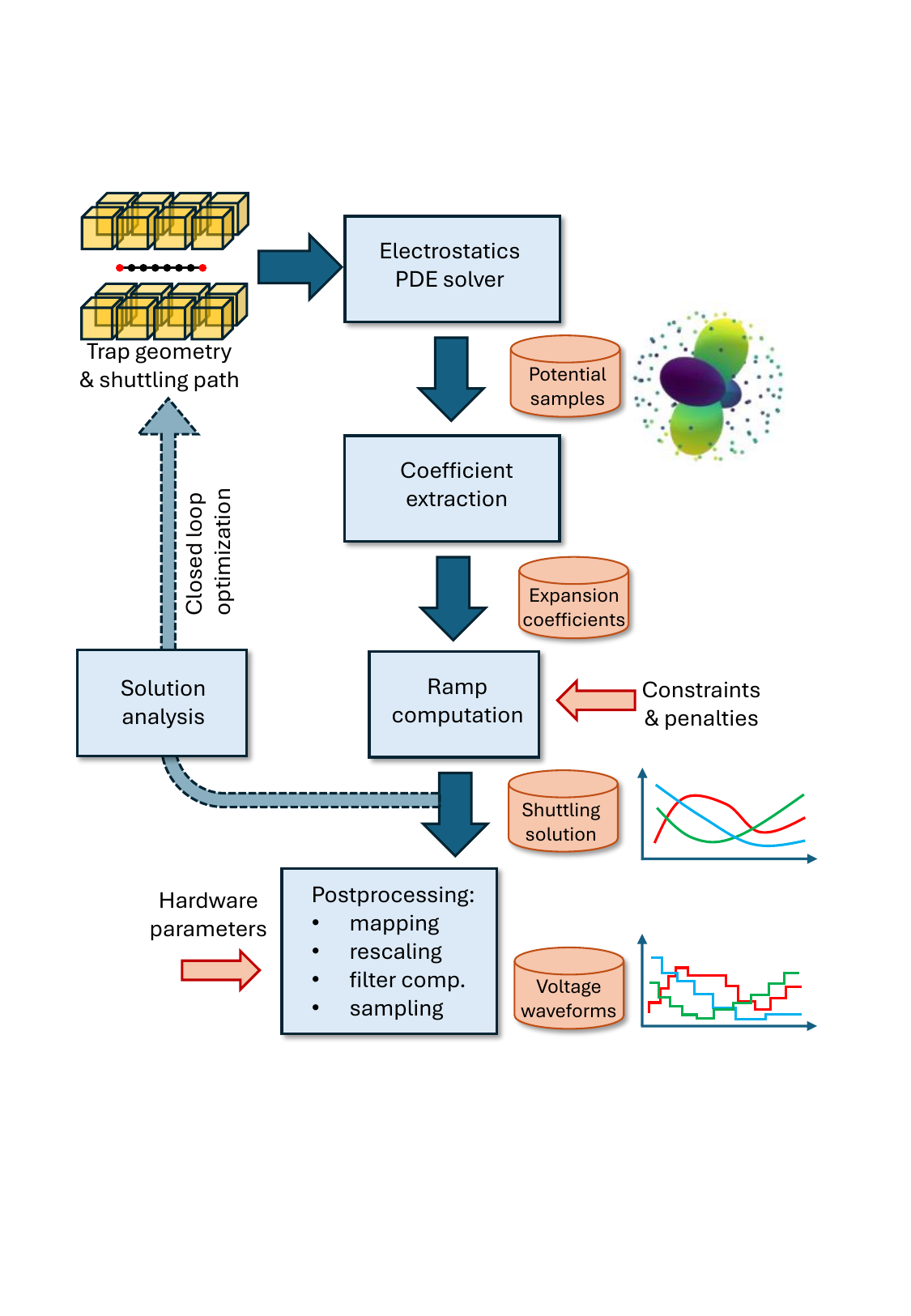}
    \caption{\justifying Schematic overview of the numerical shuttling toolchain presented in this work. An electrostatics solver is used for computing potential values near support points of a shuttling path. The trap electrode potentials are used to determine efficient multipole expansions. Based on these, optimal voltage set sequences are computed for realization of the shuttling operation. Several postprocessing steps are employed to generate voltage waveforms which can be generated a multichannel arbitrary waveform generator. Shuttling solutions can also be analyzed to e.g. perform parametric trap design studies of for closed-loop optimization of trap geometries.}
	\label{fig:toolchain_workflow}
\end{figure}

The current maturation of quantum computer platforms entails the demand of increasingly complex methods and tools for designing hardware components and control protocols for operations on the physical level. Such tools are crucial prerequisites for the realization of platforms offering genuine quantum advantage for meaningful computational tasks, and their development and deployment requires combined expertise from physics, engineering, mathematics and computer science. \\
In this work, we present a numerical toolchain \footnote{The software package developed and used in this work is available from the authors upon reasonable request.} targeting one of the most currently most mature quantum computer platforms promising near-term application-grade quantum advantage, namely platforms based on atomic ions trapped in segmented microchip radiofrequency traps. One particularly promising realization, the \emph{QCCD approach} \cite{Kielpinski2002,kaushal2020} - is based on local manipulation of small sub-registers of trapped-ion qubits via radiation, in conjunction with physical movement of the ions between different processing and storage locations in a segmented trap. Employing movement operations - termed \emph{shuttling} - one seeks to retain a sufficient degree of local control on the qubits, while enabling scalability to increasingly large register sizes. Recently, such platforms \cite{QuantinuumRaceTrack} have successfully demonstrated various quantum protocols with up to tens of qubits and high operational fidelities \citep{decross2025,ryananderson2021,Hilder2022FTR}, and architectures for scaling to much larger register sizes have been proposed \cite{lekitsch2017, Malinowski2023, valentini2025}.\\
The shuttling operations are performed by applying suitable pre-computed voltage waveforms, generated by a multichannel arbitrary waveform generator \cite{kaushal2020} to the trap electrodes. The waveforms have to be designed such that the residual motion of the ions after completion of the operation is small, as subsequent radiation-driven gate operations require localization of the ions on small length scales, for example given by optical wavelengths or tight foci of laser fields mediating the gates. On the other hand, it is desirable to execute the shuttling operations as fast as possible, in order to reduce the operational overhead \cite{Pino2021,Hilder2022FTR}. Note that while shuttling-induced  motion can in principle be removed via sympathetic cooling \cite{Pino2021}, it is still highly desirable to avoid excess motion.\\
The toolchain presented in this work serves the purpose of computing suitable voltage waveforms for enabling shuttling operations at high speed and precision, informed on the hardware resources of a given platform. The task of finding voltage waveforms for a particular shuttling operation can be formulated as follows: One seeks to determine a temporally ordered sequence of sets of electrode voltages, such that confining potential wells move within the trap in a predefined way. This task can be broken down into two mostly decoupled problems: First, one has to find voltage sets which place potential wells at given locations within the trap. Second, actual time-dependent voltage waveforms have to be generated from the voltage set sequence. The waveforms determine the dynamics of the shuttling operations and have to be generated in view of a suitable tradeoff between speed and residual motion. For this, one typically distinguishes between \emph{adiabatic} shuttling, where the voltage waveforms lead to sufficiently slow movement of the potential wells to ensure low motional excitation, and \emph{diabatic} shuttling \cite{BOWLER2012,WALTHER2012}, where control techniques are applied. A variety of such techniques exists and has been studied for the purpose of ion shuttling, such as optimal control \cite{Schulz2006}, invariant-based inverse engineering \cite{Torrontegui2011,Palmero2015} or bang-bang control \cite{Alonso2013}. These techniques are beyond the scope of this work, and we refer to \cite{Fuerst2014} for a comparative overview. However, such control strategies are usually formulated in terms of equilibrium positions and secular frequencies as the control parameters, while in practice, the accessible control parameters are the electrode voltages. Putting fast and accurate motion control protocols into practice therefore requires a numerical `gearbox' for translating between the hardware-accessible and physical-layer controls, which is precisely the toolchain presented in this work. \\
The numerical toolchain for computing shuttling solutions described in this work is sketched in Fig. \ref{fig:toolchain_workflow}. It is based on a numerical partial differential equation solver, which computes electrostatic potentials generated by the trap electrodes as solutions of Laplace's equation with Dirichlet boundary conditions \cite{Singer2010}, specified by the trap geometry and applied voltages. We employ the commercial software \textsc{NULLSPACE} \cite{nullspace} for this purpose due to its performance. We do not dwell on details on this part of the pipeline and refer to it in the following as \emph{electrostatics solver}. The actual voltage set sequence computation requires an efficient and accurate method to represent and query the electrode potentials, including the computation of confinement characteristics. In this work, we present a method for unambiguous local expansion of trap potentials into spherical harmonics, based on the concept of spherical $t$-designs. We show how higher-order derivatives of the trap potentials can be extracted, and provide an assessment of the accuracy limits of the method set. This provides the foundation for the next step, namely the computation of shuttling solutions from a suitable parametrization of the desired shuttling operation. Based on related concepts and ideas from earlier work \cite{Reichle2006,Singer2010,Nicodemus2017,Qi2021}, we establish a versatile framework expressing a shuttling problem as an unconstrained quadratic optimization problem. Here, a multi-target optimization is controlled by penalty functions in a way which allows for problem-specific and hardware-informed generation of viable shuttling solutions. The numerical efficiency of the entire pipeline is optimized to a degree of performance which allows for its direct integration into the  transpiler stack of a shuttling-based trapped-ion platform \cite{Janis2022}. Another possible application of the framework, enabled by its numerical performance, would the closed-loop optimization of trap geometry parameters, which can aid the design of electrode structures of even entire trap layouts. \\
This manuscript is structured as follows: In Sec. \ref{sec:potentials}, we describe an efficient and accurate representation of electrostatic trap potential via multipole expansions, we establish the tools for parametrizing shuttling operations and describe how local confinement properties can be computed. In Sec. \ref{sec:sitcons}, we show how the task of computing shuttling solutions can be mapped to a large linear problem, while in Sec. \ref{sec:numerics}, we discuss implementation details and features of the entire numerical pipeline and provide example data demonstrating its accuracy and performance limits. In Sec. \ref{sec:linearshuttling}, we apply the framework to a simple shuttling problem, namely shuttling between different storage segments in a linear, uniform segmented trap, before we proceed to a highly nontrivial task of shuttling in a trap junction in Sec. \ref{sec:junction}. Finally, in Sec. \ref{sec:postprocessing}, we briefly outline the basic steps which are required for computing actual dynamical voltage waveforms from a given shuttling solution.

\section{Confinement properties from spherical harmonics expansion of electrostatic potentials}
\label{sec:potentials}
We consider a segmented trap consisting of electrodes $n=1\hdots N$. Note that some scenarios, it is convenient to group physical trap electrodes into electrode sets. In the following, we always refer to \emph{electrodes}, but implicitly include this option. Applying a voltage of +1~V to electrode  $n$ and biasing all other electrodes to 0~V gives an electrostatic potential - the dimensionless \emph{unit potential} - denoted $\phi_n(\bs{r})$. The total electrostatic potential for an arbitrary applied voltage set $\{V_n\}$ is then given by linear superposition
\begin{equation}
\Phi(\bs{r})=\sum_n V_n \phi_n(\bs{r}).
\label{eq:essuperposition}
\end{equation}
The unit potentials are computed for a given trap geometry by using an electrostatics solver, which computes an influence matrix describing the electrostatic potential at a surface element, induced from a surface charges carried by any other surface element. The inversion of the influence matrix for a given trap geometry then allows for computing the surface charges on the electrode surface elements for given applied electrode voltages. Finally, with the surface charges for unit voltage settings available, the corresponding unit potentials can be computed for any point within the trap volume via Green's functions \cite{Singer2010}. We note that for the case of planar electrodes geometries, direct computation of the potentials and fields via a gapless plane approximation is also possible \cite{Wesenberg2008,Wesenberg2009}.

\begin{figure*}[!htp]
    \centering
    \includegraphics[width=0.9\textwidth,trim={0cm 5cm 2cm 4cm},clip]{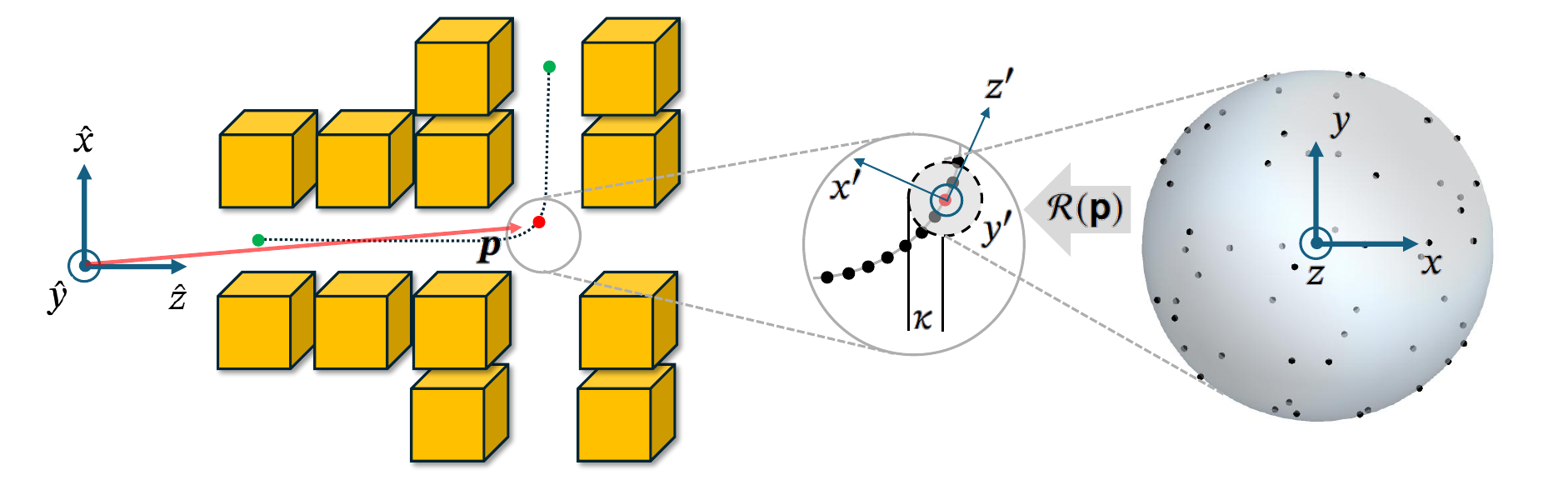}
    \caption{\justifying Trap potential evaluation: The trap geometry is defined within coordinate system $\hat{x},\hat{y},\hat{z}$, as well as a shuttling path (dotted line with green end points). A primed coordinate system is attached to each support point $\bs{p}$ along the shuttling path. Furthermore, an expansion sphere with radius $\kappa$ is set up for each support point, the surface which hosts the set of $\mc{P}$ of $K$ design points $\bs{r}_k$ defined in a coordinate system $x,y,z$ and shown on the right end. For each support point $\bs{p}$ of the shuttling path, the design points are transformed to the trap coordinate system by $\mc{R}(\bs{p})$, such that the rotated axes of the sphere coincide with the ones of the primed coordinate system centered at $\bs{p}$ (see Eq. \ref{eq:designPointsTransform}). Evaluating the electrode potentials at the transformed design points allows for an accurate and efficient representation of the trap potentials.} 
	\label{fig:sketchShuttlingPathPotentials}
\end{figure*}
Computing voltage ramps for generating viable shuttling solutions requires the capability of computing electric fields, potential curvatures and eventually higher-order derivatives of the unit potentials. The accurate computation of such derivatives can be cumbersome, as approaches based finite difference methods or fitting to model functions can lead to hardly controllable systematic errors. Furthermore, extracting derivatives from potential values computed on a three-dimensional grid can be numerically demanding: Consider a shuttling path composed of $T$ intermediate support points, and assume that for each point along the path, the unit potentials on a regular grid of $L^3$ vertices ought to be computed for extracting the required derivatives - this requires invoking an electrostatics solver on $\mathcal{O}(L^3 T N)$ points in total. We therefore emphasize that a numerical toolchain for computing shuttling solutions ultimately requires an efficient and accurate method for  computation of all required derivatives.\\
Our framework is based on the fact that the confinement properties - i.e. equilibrium positions, secular trap frequencies and orientations of the secular modes - are only of interest along predefined \emph{shuttling paths}. 
Therefore, the framework is based on a representation of the trap potentials which is illustrated in Fig. \ref{fig:sketchShuttlingPathPotentials}: For a given trap geometry and a given shuttling operation, a shuttling path comprised of a finite number of support points is defined via a suitable parametrization, and suitable representations of trap potentials are computed at each of these points. Note that a shuttling path may consist of a fixed single position, for example for a shuttling operation such as ion crystal rotation \cite{KAUFMANN2017}, throughout which the orientation of a potential ellipsoid is changed, but not its location. The shuttling paths are set up on a global trap coordinate system $\hat{x},\hat{y},\hat{z}$, on which the trap geometry is defined and the electrostatics solver operates. Then, a finite set of support points $\bs{p}_t,\; t=1\hdots T$ is placed along the path, with the index $t$ labeling \emph{sequence steps}. The shuttling operation is realized by a sequence of electrode voltage sets consisting of \emph{voltage samples} $V_{n,t}$, such that for each sequence step $t$, a combined electrostatic and ponderomotive potential ellipsoid is obtained. Ideally, the ellipsoid at sequence step $t$ is centered at $\bs{p}_t$ and has specified orientations and axes lengths, which determine the orientations and frequencies of the secular modes. A local coordinate system $\bs{\hat{x}}',\bs{\hat{y}}',\bs{\hat{z}}'$ is attached to each support point, which allows for convenient specification of the semi-axes. As exemplified in Fig. \ref{fig:sketchShuttlingPathPotentials}, arranging the primed axes to be tangential and perpendicular to a shuttling path is convenient for specifying a potential ellipsoid which remains aligned with the path throughout the shuttling operation.\\ 
In the next section, we describe how the gradients and Hessians (and eventually higher-order terms) of the unit potentials are obtained from an electrostatics solver. The determination of the confinement properties from the gradients and Hessians of the dc and rf unit potentials is discussed further below in Secs. \ref{sec:dcrfpotentials},\ref{sec:mulitpoleExpansionEandH}. 

\subsection{Multipole expansion of the unit potentials using spherical designs}
\label{sec:multipoleExpansion}
\begin{figure}[!htp]
    \centering
    \includegraphics[width=0.8\columnwidth,trim={0 0 24cm 0},clip]{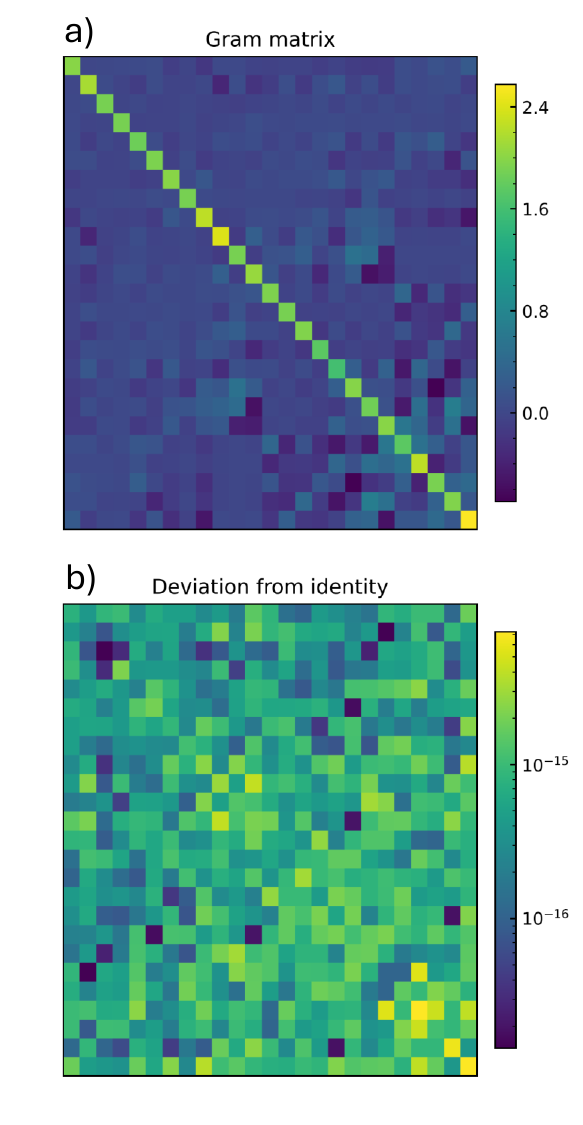}
    \caption{\justifying Reconstruction of an example unit potential $\phi=0.3\;R_{2,0}+0.7\;R_{2,2}+1.0\;R_{4,-2}$ with a Fibonacci grid of $K=25$  points. \textbf{a)} shows the Gram matrix $G$ Eq. \ref{eq:gramMatrix} prior to orthogonalization.  \textbf{b)} shows
    \label{fig:gram_matrix_plots} the difference (modulus) of the Gram matrix for the orthogonalized basis vectors and the unit matrix.}
\end{figure}

The method presented here relies on real regular solid spherical harmonics used as basis functions to expand all unit potentials around each point along the shuttling path. These are defined as
\begin{equation}
R_{l,m}(\bs{r}) = \frac{1}{r^l} \begin{cases} 
Y_{l,0}(\bs{r}) & m=0 \\ 
\tfrac{1}{\sqrt{2}}\left(Y_{l,m}(\bs{r})+(-1)^m Y_{l,-m}(\bs{r})\right) & m > 0 \\
\tfrac{1}{\sqrt{-2}}\left(Y_{l,m}(\bs{r})-(-1)^m Y_{l,-m}(\bs{r})\right) & m < 0
\end{cases}
\end{equation}
with $r=\vert\bs{r}\vert$ and the $Y_{l,m}$ being the complex-valued standard spherical harmonics. Note that similarly to the standard spherical harmonics, the regular solid spherical harmonics are orthonormal with respect to integration over the surface of a unit sphere, 
\begin{equation}
\int_{\mathbb{S}^2} R_{l,m}(1,\Omega) R_{l',m'}(1,\Omega) d\Omega = \delta_{ll'}\delta_{mm'}
\label{eq:RfnOrthogonality}
\end{equation}
and are also solutions of the Laplace equation, i.e. $\Delta R_{l,m} = 0$. Note that for our purposes, the $R_{l,m}(\bs{r})$ need to be explicitly specified in Cartesian coordinates. We require an as-small-as possible set of $K$ points arranged on the surface of the unit sphere:
\begin{equation} 
\mc{P}=\{\bs{r}_1,\hdots,\bs{r}_K \in \mathbb{S}^2 \},
\label{eq:setOfDesignPoints}
\end{equation}
on which we evaluate the unit potential of interest. From the resulting potential values, we seek to obtain an unambiguous expansion of the unit potential around a given support point. An expansion up to order $L$ comprises $(L+1)^2$ basis functions. The point set $\mc{P}$ would be an exact spherical $t$-design with $t=2 L$ if the orthogonality relation Eq. \ref{eq:RfnOrthogonality} is preserved upon evaluation on the finite point set:
\begin{equation}
    \sum_{k=1}^K R_{l,m}(\bs{r}_k) R_{l',m'}(\bs{r}_k) \propto \delta_{ll'}\delta_{mm'}
\label{eq:sphericaltDesign}
\end{equation}
which is a sufficient condition for reliable extraction of the expansion coefficients from the potential values evaluated on point set $\mc{P}$. Generating spherical $t$-designs can be cumbersome, furthermore less points than contained in an exact $t$-design may actually be sufficient to achieve good accuracy for a given expansion order $L$. Therefore, we use spherical Fibonacci grids to generate point sets $\mathcal{P}$ with good uniform and equidistant distribution on the unit sphere, see Appendix \ref{sec:AppFibonacci} for details.\\
We evaluate the spherical basis functions at all points of the set $\mc{P}$:
\begin{equation}
\bs{R}_{l,m}=\left( R_{l,m}(\bs{r}_1),\hdots,R_{l,m}(\bs{r}_K)\right)^T,
\end{equation}
and arrange these in a $(L+1)^2 \times K$ matrix $V$ comprised of $(L+1)^2$ $K$-dimensional row vectors $\bs{R}_{l,m}$, or respectively of $K$ $(L+1)^2$-dimensional column vectors $\bs{v}_i$:
\begin{equation}
V=\begin{pmatrix} \bs{R}_{0,0}^T  \\ \ \bs{R}_{1,-1}^T  \\ \vdots \\ \bs{R}_{4,4}^T   \end{pmatrix}=\left( \bs{v}_1, \hdots, \bs{v}_{(L+1)^2} \right).
\end{equation}
With $\mc{P}$ failing to be a proper spherical 8-design, Eq. \ref{eq:sphericaltDesign} does not hold and consequently, the column vectors are not orthogonal:
\begin{equation}
    \bs{v}_i \cdot \bs{v}_j \not\propto \delta_{ij}.
\end{equation}
From the Gram matrix
\begin{equation}
G=\{g_{ij}\}=\{\bs{v}_i \cdot \bs{v}_j\},
\label{eq:gramMatrix}
\end{equation}
we obtain a transformed matrix
\begin{equation}
U=G^{-1/2} V= \left( \bs{u}_1, \hdots,  \bs{u}_{(L+1)^2} \right),
\label{eq:basisVectorsOrthogonalized}
\end{equation}
which is comprised of orthonormal column vectors:
\begin{equation}
    \bs{u}_i \cdot \bs{u}_j = \delta_{ij}.
    \label{eq:gramMatrixOrthogonalized}
\end{equation}
We can use these to expand the unit potential $\phi$ for any electrode (or electrode set) at any point $\bs{p}$, specified in the trap coordinate system. Let $\mathcal{R}(\bs{p})$ be the $SO(3)$ rotation matrix which rotates the global coordinate axes $\bs{\hat{x}},\bs{\hat{y}},\bs{\hat{z}}$ onto local axes $\bs{\hat{x}'},\bs{\hat{y}'},\bs{\hat{z}'}$ and $\kappa$ a suitably chosen scaling parameter (typically a small fraction of a reference trap geometry parameter). We apply a Poincar\'e transform to the design points from set $\mathcal{P}$ Eq. \ref{eq:setOfDesignPoints} to obtain a set $ \{\bs{q}_1,\hdots,\bs{q}_K\}$ of $K$ points in the trap coordinate system, distributed on the surface of a sphere of radius $\kappa$ and centered at $\bs{p}$:
\begin{equation}
\bs{q}_k= \bs{p}+\kappa \;\mathcal{R}(\bs{p})\; \bs{r}_k.
\label{eq:designPointsTransform}
\end{equation}
Using the electrostatics solver, we evaluate the unit potential of interest at these points and arrange the results in a $K$-element vector $\bs{\phi}$:
\begin{equation}
\bs{\phi} =\{\phi(\bs{q}_1),\hdots,\phi(\bs{q}_K)\}.
\end{equation}
We can now directly project out the components of $\bs{\phi}_n$ in the orthonormal basis given by the column vectors of $U$, obtaining an $(L+1)^2$-dimensional column vector of expansion coefficients:
\begin{equation}
    \bs{\tilde{c}}=(\tilde{c}_{i})^T = (\bs{\phi} \cdot \bs{u}_i)^T.
\end{equation}
In the next step, we seek to obtain expansion coefficients uniquely and unambiguously pertaining to the spherical basis functions. To that end, we revert the orthonormalization to  obtain a new coefficient vector from the coefficient vector $\bs{\tilde{c}}$:
\begin{equation}
\bs{c} = \left(G^{1/2}\right)^T \bs{\tilde{c}}.
\label{eq:orthogonalizedCoeffs}
\end{equation}
Finally, we have to rescale the coefficients with the power of the sphere radius $\kappa$, pertaining to the respective value of $l$:
\begin{equation}
c_i \leftarrow c_i / \kappa^{l_i}.
\end{equation}
As a result, we obtain an expansion of $\phi(\bs{r})$ in the spherical basis around $\bs{p}$ with respect to the primed axis system:
\begin{equation}
\phi(\bs{r}') = \sum_{i=1}^{(L+1)^2} c_i\; R_{l_i,m_i}(\bs{r}').
\end{equation}
Fig. \ref{fig:gram_matrix_plots} shows the original Gram matrix Eq. \ref{eq:gramMatrix} and the deviation of orthogonalized Gram matrix Eq. \ref{eq:gramMatrixOrthogonalized} from the identity, for a harmonic example potential. It can be seen that the difference of Gram matrix for the orthogonalized basis vectors and the unit matrix exhibits numerical deviations below the 10$^{-14}$ level. The reconstructed coefficients $c_i$ Eq. \ref{eq:orthogonalizedCoeffs} (not shown) match one of the input potential up to the same accuracy, which illustrates that the a posteriori orthogonalization boosts the accuracy of the multipole expansion. From the resulting multipole expansion of the unit potentials, all derivatives of up to order $L$ can be computed without introducing additional numerical errors. 

\subsection{Electrostatic, ponderomotive and total trap potentials}
\label{sec:dcrfpotentials}
In this section, we establish how to generally characterize confinement in terms of residual forces and Hessians for combined electrostatic and ponderomotive forces and general trap geometries. Here and in the following, we denote all quantities derived from unit-voltage potentials with small letters ($\phi,\bs{e},h$) and quantities scaled with applied trap voltages with capital letters ($\Phi,\bs{E},H$). Implicitly, all these quantities are evaluated at a support point $\bs{p}$, to which a local coordinate system ($x',y',z'$) with basis vectors  $\bs{\epsilon}_s$ is attached, see Fig. \ref{fig:sketchShuttlingPathPotentials}.\\
The unit voltage electric field (in units of $m^{-1}$) generated by electrode (set) $n$ and expressed in a local coordinate system, reads
\begin{equation}
\bs{e}_n=-\boldsymbol{\nabla}'\; \phi_n = -\sum_s \partial_s \phi_n \;\boldsymbol{\epsilon}_s.
\label{eq:En} 
\end{equation}
The unit voltage Hessian matrix (in units of $\text{m}^{-2}$) containing all second derivatives of the unit potential for electrode $n$, is given by
\begin{equation}
h_n = \left(\bs{\nabla}'\otimes \bs{\nabla}'\right)\phi_n = \sum_{r,s} \partial_r \partial_s \phi_n \;\bs{\epsilon}_r \otimes \bs{\epsilon}_s.
\label{eq:Hn}
\end{equation}
The total electrostatic field and Hessian at support point $\bs{p}$ with electrode voltages $\{V_n\}$ applied are given by
\begin{equation}
    \bs{E}_{\text{dc}}=\sum_n V_n \bs{e}_n \qquad H_{\text{dc}}=\sum_n V_n h_n.
    \label{eq:dcTotalFieldandH}
\end{equation}
Confinement in a radiofrequency (rf) trap requires at least one additional oscillatory electric field, which generates a time-averaged \emph{ponderomotive} potential (or pseudo-potential). We consider a trap with rf electrodes supplied with a single rf signal at fixed voltage amplitude $V_{\text{rf}}$ and drive frequency $\Omega$, which generates the electrostatic unit potential $\phi_{\text{rf}}$, the unit electric field $\bs{e}_{\text{rf}}$ and the unit Hessian $h_{\text{rf}}$. The resulting ponderomotive potential for a particle with charge $Q$ and mass $m$ in an electric field $\bs{\mathcal{E}}_{\text{rf}}=V_{\text{rf}}\bs{e}_{\text{rf}}$ oscillating at frequency $\Omega$, expressed in Volts for convenience, is given by:
\begin{equation}
\Phi_{\text{rf}}=\frac{Q\;V_{\text{rf}}^2}{4m\Omega^2}  \left(\bs{e}_{\text{rf}} \cdot \bs{e}_{\text{rf}}\right) =\frac{\alpha_{\text{rf}}}{2}   \sum_s   \left(\partial_s \phi_{\text{rf}}\right)^2, 
\label{eq:ponderomotivePotential}
\end{equation}
where we have lumped constants into the prefactor (units of $Vm^2$)
\begin{equation}
\alpha_{\text{rf}}=\frac{Q\;V_{\text{rf}}^2}{2m\Omega^2}. 
\end{equation}
Note that we specify ponderomotive forces as \emph{effective electric fields}, which is convenient for treating these on the same footing as electrostatic forces. An inhomogeneous oscillating field gives rise to a ponderomotive force, expressed as an effective electric field (in units of $V/m$):
\begin{eqnarray}
\bs{E}_{\text{rf}}&=&-\bs{\nabla}' \Phi_{\text{rf}} \nonumber \\
&=&-\alpha_{\text{rf}}  \sum_{r,s} \bs{\epsilon}_r \; \partial_s \phi_{\text{rf}}\; \partial_r\partial_s \phi_{\text{rf}}  \nonumber \\
&=& \alpha_{\text{rf}} h_{\text{rf}} \bs{e}_{\text{rf}}.
\label{eq:rfforce}
\end{eqnarray}
The last equality is obtained by using Eqs. \ref{eq:En},\ref{eq:Hn} for the rf unit potential and $\left(\bs{\epsilon}_r \otimes \bs{\epsilon}_s\right) \bs{\epsilon}_{s'}=\left(\bs{\epsilon}_s\cdot \bs{\epsilon}_{s'}\right) \bs{\epsilon}_r=\delta_{ss'}\bs{\epsilon}_r$. \\
The Hessian for the pseudopotential (in units of $V/m^2$) is
\begin{eqnarray}
H_{\text{rf}} &=&  \bs{\nabla}'\otimes \bs{\nabla}'\Phi_{rf} \nonumber \\ 
&=& \frac{\alpha_{\text{rf}}}{2} \sum_{r,s,s'}  \bs{\epsilon}_r \otimes \bs{\epsilon}_s\; \partial_r\partial_s \left(\partial_{s'} \phi_{\text{rf}}\right)^2 \nonumber \\
&=& \alpha_{\text{rf}} \smash{\underbrace{h_{\text{rf}}^2}_{\displaystyle \tilde{h}_{\text{rf}}^{(a)}}}+\alpha_{\text{rf}}\smash{\underbrace{(\bs{e}_{\text{rf}}\cdot\bs{\nabla}')h_{\text{rf}}}_{\displaystyle  \tilde{h}_{\text{rf}}^{(b)}}}  \label{eq:Hrf}. \\
\nonumber \\
\nonumber
\end{eqnarray}
The term $\tilde{h}_{\text{rf}}^{(b)}$ vanishes in the rf-null, i.e. at positions where $\bs{e}_{\text{rf}}=0$. \\

Our framework rests on the assumption that electrostatic and ponderomotive potentials as well as the forces resulting from these can simply be added. A necessary criterion is separation between secular and drive frequencies $\omega_u \ll \Omega$, corresponding to trap operation deep in the stability region. Additional criteria allowing for superposition of static and ponderomotive forces are listed in Appendix \ref{sec:rfdcValidity}. \\
The total effective electric field (in V/m) is given by the sum of the electrostatic and ponderomotive fields: 
\begin{equation}
\bs{E}=\bs{E}_{\text{rf}}+\bs{E}_{\text{dc}} = \alpha_{\text{rf}} h_{\text{rf}} \bs{e}_{\text{rf}}+ \sum_n V_n \bs{e}_n.
\label{eq:totalforce}
\end{equation}
The combined Hessian pertaining to all dc and rf electrodes (in $V/m^2$) is
\begin{equation}
H=H_{\text{rf}}+H_{\text{dc}}= \alpha_{\text{rf}} \left(\tilde{h}_{\text{rf}}^{(a)}+\tilde{h}_{\text{rf}}^{(b)}\right)+ \sum_n V_n h_n. 
\label{eq:Htot}
\end{equation}
It holds that $\bs{E}=0$ if a potential well is located at position ${\bs{p}}$, around which the total mechanical potential $U=Q\Phi$ varies up to second order along direction $\bs{\xi}=r\bs{\hat{\xi}}$ as:
\begin{equation}
U\approx \frac{1}{2} Q \bs{\hat{\xi}} H \bs{\hat{\xi}}\; r^2.
\label{eq:Htotquadraticform}
\end{equation}
If the unit vector $\bs{\hat{\xi}}$ happens to be an eigenvector of $H$ to eigenvalue $\lambda_u$, the secular frequency pertaining to that direction is given by
\begin{equation}
     \omega_u^2 = \lambda_u\frac{Q}{m}.
    \label{eq:secularFreqFromH}
\end{equation}
The formalism presented above is chosen to meet the demands of the framework, but deviates from standard treatments of radiofrequency traps. We e stablish the link to the latter by considering a simplified scenario with only one dc electrode and coinciding symmetry centers of the dc and rf potentials. We express both dc and rf potentials in terms of curvature parameters  $c_{\text{dc}}, c_{\text{rf}}$ and an effective electrode dimension $\tilde{r}$:
\begin{equation}
\phi_{\text{dc,rf}}(\bs{r})=\sum_{u=x,y,z}c_{\text{dc,rf}}^{(u)} \frac{u^2}{\tilde{r}^2}. 
\label{eq:standardPotentials1D}
\end{equation}
Computing the respective second derivatives, the second derivative of the ponderomotive potential Eq. \ref{eq:ponderomotivePotential} according to 
 Eq. \ref{eq:Hrf}, and the second derivative of the total potential from Eq. \ref{eq:Htot}, the secular frequency along direction $u$ from Eq. \ref{eq:secularFreqFromH} is given by the familiar form
\begin{equation}
\omega^2=\frac{\Omega^2}{4}\left(a+\frac{q^2}{2}\right),
\end{equation}
with the usual Mathieu coefficients $a,q$
\begin{equation}
    a=\frac{8QV_{\text{dc}}c_{\text{dc}}^{(u)}}{m\tilde{r}^2\Omega^2} \qquad q=\frac{4QV_{\text{rf}}c_{\text{rf}}^{(u)}}{m\tilde{r}^2\Omega^2}.
\end{equation}

\subsection{Forces and curvatures from multipole expansion}
\label{sec:mulitpoleExpansionEandH}
Once the spherical expansion coefficients are obtained for all relevant electrodes at a given position $\bs{p}$ and with respect to the chosen axes system, the total forces and potential curvatures can be computed. 
To that end, we compute the derivatives of the basis functions $R_i$ at the origin of their proper coordinate system $x,y,z$ (see Fig. \ref{fig:sketchShuttlingPathPotentials}), which leads to constant coefficients not depending on $\bs{p}$. Weighing these coefficients with the expansion coefficients computed in the primed system attached to $\bs{p}$ then gives the required derivatives for these coordinates. From computing forces and curvatures, including the ponderomotive potential, an expansion order of $L=3$ is sufficient. For points $\bs{p}$ restricted to the rf-null, an expansion order of $L=2$ is sufficient. In the following, all expansions coefficients $c_i$  implicitly depend on the electrode index $n$ and the support point $\bs{p}$.\\
For computing the unit voltage electric field Eq. \ref{eq:En} for an electrode of interest, only $l=1$ components need to be taken into account:
\begin{eqnarray}
\bs{e}_n &=& -\sum_{i=2}^4 c_i\; \bs{\nabla}\; R_i(\bs{r}) \vert_{\bs{r}=0} \nonumber \\
&=& \sqrt{\frac{3}{4\pi}}\left(c_4,  c_2, -c_3 \right)^T.
\label{eq:enFromCoeffs}
\end{eqnarray}
For computing the unit voltage Hessians, only the five coefficients pertaining to $l=2$ need to be taken into account:
\begin{equation}
h_{r,s}^{(n)} = \sum_{i=5}^9 c_i\; \partial_{r}\partial_{s}\, R_i(\bs{r}) \vert_{\bs{r}=0}
\end{equation}
which gives
\begin{equation}
h^{(n)} = -\sqrt{\frac{15}{4\pi}} \begin{pmatrix} \tfrac{1}{\sqrt{3}} c_7 - c_9& c_5 & c_8 \\
c_5 & \tfrac{1}{\sqrt{3}} c_7 + c_9 & c_6 \\
c_8 & c_6 & -\tfrac{2}{\sqrt{3}} c_7
\end{pmatrix}.
\label{eq:hnFromCoeffs}
\end{equation}
Note that the rf electrode unit voltage fields $\bs{e}_{\text{rf}}$ and Hessians $h_{\text{rf}}$ have the same expansion coefficient representations as the corresponding dc quantities.\\ 
The ponderomotive force Eq. \ref{eq:rfforce} computes from the expansion coefficients as
\begin{equation}
\bs{E}_{\text{rf}}=\frac{3\sqrt{5}}{4\pi}\alpha_{\text{rf}}\begin{pmatrix} 
c_3c_5-\tfrac{1}{\sqrt{3}}c_2c_7 -c_4c_8 \\
-c_2c_5-c_4c_6+\tfrac{1}{\sqrt{3}}c_3c_7+c_3c_9 \\
c_3 c_6+ \tfrac{2}{\sqrt{3}} c_4 c_7 -c_2c_8
\end{pmatrix}.
\label{eq:ErfFromCoeffs}
\end{equation}
Explicit expressions for both parts of the pseudopotential Hessian $H_{\text{rf}}$ Eq. \ref{eq:Hrf} in terms of the expansion coefficients are given in Appendix (Eqs. \ref{eq:htildea}, \ref{eq:htildeb}).

\section{Computing shuttling solutions}
\label{sec:sitcons}
Given the established capability to accurately and efficiently compute derivatives of the unit potentials for all trap electrodes at any given position within the trap volume, we now describe how these can be used to compute sequences of voltage sets - termed \emph{shuttling solutions} - providing the basis for realizing a given shuttling operation. This means that each sequence step $t=1\hdots T$, we seek to establish $w=1\hdots W$ \emph{potential wells} centered at positions $\boldsymbol{r}_{w,t}$ specified in the trap coordinate system. For any given well $w$, the set of well positions $\{\boldsymbol{r}_{w,t}\}$ defines a \emph{shuttling path} (see Fig. \ref{fig:sketchShuttlingPathPotentials}). Note that $t$ and $T$ do not specify actual timings, but nevertheless refer to temporally ordered sequences. Throughout a shuttling process, two well positions can emerge from (or converge to) the same location, which realizes splitting (or merging) operations.\\ 
We emphasize at this point that a shuttling solution is just an ordered sequence of electrode voltage sets suitable for quasi-static realization of a shuttling operation, it does neither take actual physical dynamics into account, nor do the steps $t$ correspond to actual timings. The subsequent computation of actual \emph{voltage waveforms} based on concrete hardware parameters and ion dynamics is covered further below in Sec. \ref{sec:postprocessing}.\\ 
A potential well being characterized by a confining potential ellipsoid, we also specify the orientation of its main axis and the curvatures along these. The task now consists of determining voltage levels $V_{n,t}$ for each dc electrode $n$ and sequence step $t$, which realize the desired shuttling operation as good as possible, informed on limited resources such as bandwidth and voltage range. In the following, we express the task of determining optimal voltage set sequences for a given shuttling process as an unconstrained quadratic optimization problem, consisting of several penalty terms $\mc{F}^{(p)}$. Such problem is generally equivalent to solving a linear system, which allows for straightforward and efficient numerical solution, without resorting to numerical optimization procedures.\\
\subsection{Shuttling penalties}
Each penalty consists of a penalty kernel $\mc {F}^{(p)}_{u,t,w}$ and a weight vector $ \mc{W}_{u,t,w}^{(p)}$, and the products of these are summed over the sequence step $t$, wells $w$ and spatial directions $u={x',y',z'}$:
\begin{equation}
    \mc{F}^{(p)} = \sum_{u,t,w} \mc{W}_{u,t,w}^{(p)} \mc {F}^{(p)}_{u,t,w}.
\end{equation}
The weight factors $\mathcal{W}_{u,t,w}^{(p)}$ serve to set the relative importance of the penalty, and may depend on $u,t,w$. The quality of the optimization results may strongly depend on the management of the penalty terms via the weight factors, which is facilitated by separating the weights into a constant scaling factor and dimensionless factors which may depend on $u,t,w$:
\begin{equation}
 \mc{W}_{u,t,w}^{(p)} = \mc{W}_{0}^{(p)} W_{u,t,w}^{(p)}.
 \label{eq:penaltyScalingFactors}
\end{equation}
For determining suitable penalty scaling factors, we need to fix reference secular frequencies $\omega_{\text{ref},u}$, which can be the preset secular frequencies at the beginning of the shuttling operation.\\
In the following, we individually discuss the different relevant penalty terms. 

\subsubsection{Well position penalty}
For establishing a confining potential well at position $\bs{r}_{w,t}$, the total force evaluated at $\bs{r}_{w,t}$ 
\begin{equation}
    \bs{E}_{w,t}=\bs{E}(\bs{r}_{w,t}),
\end{equation}
has to vanish:
\begin{equation}
    \vert\vert\bs{E}_{w,t}\vert\vert^2 \overset{\text{!}}{=} 0.
\end{equation}
We define the corresponding penalty kernel to be the weighted sum of the squared residual force components along directions $u$: 
\begin{equation}
\mc{F}^{(1)}=\sum_{u,w,t} \mc{W}^{(1)}_{u,w,t} E_{u,w,t}^2.   
\label{eq:penalty1}
\end{equation}
where residual force components (specified as an effective electric field in $V/m$) are obtained from Eq. \ref{eq:totalforce}. This penalty serves for positioning minima of the \emph{combined} electrostatic and ponderomotive potential. Unless enforced by symmetry or appropriate grouping of trap electrodes, the minima may not coincide, leading to excess micromotion, i.e. rapid oscillations of the ion driven by a nonzero rf field. Throughout shuttling operations, transient excess micromotion may be tolerated, as it does not necessarily lead to residual motion upon completion of the shuttling. For shuttling solutions with potential wells moving outside the rf null, the criteria listed in Appendix \ref{sec:rfdcValidity} should be tested a posteriori to check the validity of the pseudopotential approximation. Furthermore, noise cross-coupling from the rf field may lead to increased heating outside of the rf null \cite{Blakestad2009,Blakestad2011,taniguchi2025}, the effect of which may be estimated once the relevant noise spectral densities are known.\\
We pin down a reference scale for controlling the penalty weight by fixing reference position deviations $\Delta u$ and computing the magnitude the penalty would assume for this deviation at a preset secular frequency $\omega_{\text{ref},u}$:
\begin{eqnarray}
QE_u&=&-m\omega_{\text{ref},u}^2 \Delta u \nonumber \\
\Rightarrow \mc{F}^{(1)}_{u,w,t}&=&\frac{m^2}{Q^2}\omega_{\text{ref},u}^4 \Delta u^2.
\end{eqnarray}
Now, if we choose 
\begin{equation}
\mc{W}^{(1)}_{0}=\frac{Q^2}{m^2 \omega_{\text{ref},u}^4 \Delta u^2},
\end{equation}
we get one unit of penalty 1 if at step $t$, well $w$ is off by $\Delta u$ along direction $u$ at secular frequency $\omega_{\text{ref},u}$.\\

\subsubsection{Confinement penalty}
It is generally beneficial to keep secular frequencies constant throughout a shuttling operation, or at least to maintain as much confinement as possible, in order to suppress excitation from inertia and squeezing. Moreover, background-noise induced heating can drastically increase at low secular frequencies \cite{Brownutt2015}. We therefore seek to fix the secular frequencies $\omega_u$ and the orientation of the main axes of the confining potential at all well positions $\bs{r}_{w,t}$, by pinning down the entries of the total Hessian Eq. \ref{eq:Htot}, evaluated at each $\bs{r}_{w,t}$:
\begin{equation}
    H_{w,t}=H(\bs{r}_{w,t}).
\end{equation}
The corresponding penalty, for each time step $t$ and well $w$, is then given by the squared Frobenius norm of the difference of the actual total Hessian $H$ and the desired one $H^{(set)}$:
\begin{eqnarray}
  \mc{F}^{(2)} &=&\sum_{\{uu'\},w,t} \mc{W}^{(2)}_{u,u',w,t} \vert\vert H_{w,t}-H_{w,t}^{(set)}\vert\vert^2_F.
  \label{eq:penalty2}
\end{eqnarray}
It can be advantageous to orient the primed coordinate system used for computing the potential coefficients at each desired well position along the desired axes (see Fig. \ref{fig:sketchShuttlingPathPotentials}). Upon choosing desired secular frequencies $\omega_{\text{ref},u}$, the desired total Hessian will then assume diagonal form:
\begin{equation}
    H^{(set)}_{u,u'} = \frac{m}{Q}\omega_{\text{ref},u}^2\delta_{uu'}.
    \label{eq:penaltyDesiredHess}
\end{equation}
Note that depending on the shuttling operation of interest, the preset secular frequencies $\omega_{\text{ref},u}$ may (but must not) depend on $w,t$. \\
For determining a suitable scaling factor for the weights, we assume diagonal actual and desired Hessians. The penalty kernel from Eq. \ref{eq:penalty2} simplifies to
\begin{eqnarray}
    H_{w,t}-H_{w,t}^{(set)} &=& \frac{m}{Q}\left(\omega_u^2-\omega_{\text{ref},u}^2\right) \delta_{uu'} \nonumber \\
    &\approx & 2\frac{m}{Q} \omega_{\text{ref},u} \Delta\omega_u.
\end{eqnarray}
Fixing reference deviations $\Delta\omega_u$ between actual and desired secular frequencies, and setting the penalty scaling factor to
\begin{equation}
    \mc{W}^{(2)}_{0}=\frac{Q^2}{4 m^2\omega_{\text{ref},u}^2 \Delta\omega^2 }
\end{equation}
yields one unit of penalty 2 when the secular frequency along $u$ is deviates from the reference frequency by $\pm\Delta\omega$.\\

\subsubsection{Voltage penalty}
The moduli of the voltage samples are generally to kept small in order to comply with the voltage limits of the employed waveform generator. The corresponding penalty reads
\begin{equation}
    \mc{F}^{(3)}=\sum_{n',t'} \mc{W}^{(3)}_{n,t}V_{n,t}^2.
  \label{eq:penalty3}
\end{equation}
The trap unit potentials are generally of long-range type. However, it is desirable to preclude that trap electrodes far away from a desired well position $\bs{r}_{w,t}$ are taken into account for the generation of the confining potential. The penalty vector $W^{(3)}_{n,t}$ can be conveniently used for this purpose by including an \emph{activation function}: Let 
\begin{equation}
D_{n,w,t}=||\bs{r}_{w,t}-\bs{R}_n||
\end{equation}
be a suitable measure of distance between a well position with $\bs{R}_n$ being a suitable specification of the location of electrode (set) $n$ within the trap coordinate system. Then, a `bathtub' type activation function
\begin{equation}
    \mc{W}^{(3)}_{n,t}=\mc{W}^{(3)}_{0}\min_{w} \begin{cases}\mathcal{L} \quad & D_{n,w,t} \geq \mathcal{D}_2 \\
\mathcal{L} \frac{D_{n,w,t}-\mathcal{D}_1}{\mathcal{D}_2-\mathcal{D}_1}\quad  & \mathcal{D}_1 \leq D_{n,w,t} < \mathcal{D}_2\\
1 \quad & D_{n,w,t}<\mathcal{D}_1, \end{cases}
\label{eq:activationFunctions}
\end{equation}
with a very large value $\mathcal{L} \gg 1$ and distances $\mathcal{D}_2 < \mathcal{D}_1$ gradually activates segment $n$ if at least one well is to be positioned in the vicinity of segment $n$.

\subsubsection{Voltage difference penalty}
The moduli of the voltage samples are supposed to be as small as possible to comply with hardware resources, see Sec. \ref{sec:filterInversion} further below. The corresponding penalty does not depend on $u,w$ and reads
\begin{eqnarray}
    \mc{F}^{(4)}&=&\sum_{n}\sum_{t=2}^S \mc{W}^{(4)}    \left(V_{n,t}-V_{n,t-1} \right)^2.
  \label{eq:penalty4}
\end{eqnarray}
As this penalty corresponds to a global hardware-imposed bandwidth limitation, we may take $\mc{W}^{(4)}$ to be independent of $n,u,t,w$. 

\subsubsection{Fixed voltage set penalty}
For some practically relevant situations, it may be required to employ fixed (calibrated or separately precomputed) voltage configurations. After terminating a shuttling operation, one might typically seek to obtain a confinement situation with precisely calibrated properties for driving sensitive quantum gate operations. Another example is the precise tuning of the critical point of vanishing harmonic confinement for a separate or merge operation \cite{KAUFMANN2014,RUSTER2014}. For simplicity, we consider a single such voltage set $\{\hat{V}_n\}$. The penalty reads
\begin{equation}
    \mc{F}^{(5)}=\sum_{n,t} \mc{W}^{(5)}_{n,t} \left(V_{n,t}-\hat{V}_{n}\right)^2.
  \label{eq:penalty5}
\end{equation}
For enforcing the reference voltage set at time step $\hat{t}$, the penalty weights need to assume large nonzero values only for $\hat{t}$. The penalty term can also straightforwardly be extended to allow for distinct voltage sets applied at different sequence steps. \\
Note that for $\hat{V}_n=0$, this penalty becomes identical to the voltage penalty Eq. \ref{eq:penalty3}, such that by proper choice the weight factors, both penalties could be merged. However, as both penalties serve different purposes, is is beneficial in terms of usability and extensibility to keep them separated in the numerical framework.

\subsection{Computing optimal solutions}
\label{sec:linearSystem}
An optimum voltage ramp is generated by finding a stationary point of the total penalty function:
\begin{equation}
    \sum_p \partial_{V_{n,t}} \mc{F}^{(p)} \overset{\text{!}}{=} 0.
\label{eq:allPenaltiesStationary}
\end{equation}
We bundle all voltage samples into a $NT$-component vector $\bs{v}$ 
\begin{equation}
V_{n,t}=\bs{v}_{(n,t)} \qquad (n,t)=N(t-1)+n.   
\label{eq:Vntindexing}
\end{equation}
Setting up a $NT\times NT$ matrix $A$ and a vector $\bs{b}$ with entries
\begin{eqnarray}
    A_{(n,t),(n',t')}&=& \sum_p \partial_{v_{(n',t')}} \partial_{v_{(n,t)}} \mc{F}^{(p)} \nonumber \\
    b_{(n,t)}&=&\sum_{n',t'} A_{(n,t),(n',t')}-\sum_p  \partial_{v_{(n,t)}} \mc{F}^{(p)} 
\end{eqnarray}
allows for casting Eq. \ref{eq:allPenaltiesStationary} into a linear system:
\begin{equation}
    A\bs{v}=\bs{b},
\label{eq:linearSystem}
\end{equation}
the solution of which provides optimum voltage sets for the shuttling task. The indexing convention Eq. \ref{eq:Vntindexing} is useful as it renders $A$ to be band-diagonal. The voltage and voltage difference penalties can be used to enforce solutions which are compatible with available hardware resources, but also render $A$ to be well-conditioned if the corresponding penalty scaling factors are sufficiently large. 
We now introduce shorthand notations such as e.g.
\begin{equation}
    e_{u,n}^{(w,t)}=e_{n,u}(\bs{r}_{w,t}),
\end{equation}
and similarly for all fields and Hessians, and conclude this section by providing explicit expressions for the entries of $A$ and $\bs{b}$:
\begin{widetext}
\begin{eqnarray}
    A_{(n,t),(n',t')}&\overset{\text{Eq.\ref{eq:penalty1}}}{=}&\sum_{u,w} \mc{W}^{(1)}_{u,w,t} e_{u,n}^{(w,t)} e_{u,n'}^{(w,t)} \overset{\text{Eq. \ref{eq:penalty2}}}{+} \sum_{u,u',w} \mc{W}^{(2)}_{u,u',w,t}  h_{u,u',n}^{(w,t)} h_{u,u',n'}^{(w,t)} \nonumber  \\
    &\overset{\text{Eq. \ref{eq:penalty3}}}{+}& \mc{W}^{(3)}_{n,t} \delta_{nn'}\delta_{tt'} \overset{\text{Eq. \ref{eq:penalty4}}}{+} \mc{W}^{(4)} \delta_{nn'} \times \begin{cases}\delta_{t',1}-\delta_{t',2} \quad \text{for} \quad & t=1 \\
\delta_{t',S}-\delta_{t',S-1} \quad \text{for} \quad & t=S   \\
2\delta_{t',t}-\delta_{t',t-1}-\delta_{t',t+1} \quad & \text{otherwise} \end{cases}
\nonumber \\
    &\overset{\text{Eq. \ref{eq:penalty5}}}{+}& \mc{W}^{(5)}_{n,t} \delta_{nn'}\delta_{tt'},
    \label{eq:AmatrixFinal}
\end{eqnarray}

\begin{eqnarray}
b_{(n,t)} &\overset{\text{Eq. \ref{eq:penalty1}}}{=}& -\sum_{u,w,t} \mc{W}^{(1)}_{u,w,t} e_{u,n}^{(w,t)} E_{\text{rf},u}^{(w,t)} \nonumber \\
&\overset{\text{Eq. \ref{eq:penalty2}}}{-}& \sum_{u,u',w} \mc{W}^{(2)}_{u,u'} h_{u,u',n}^{(w,t)} \left(H_{\text{rf},u,u'}^{(w,t)}-H^{(set,w,t)}_{u,u'}\right) \overset{\text{Eq. \ref{eq:penalty5}}}{+} \mc{W}^{(5)}_{n,t} \hat{V}_n.
    \label{eq:bvectorFinal}
\end{eqnarray}
\end{widetext}

\section{Numerical pipeline}
\label{sec:numerics}

In this section, we describe the components of our numerical framework and assess the accuracy of representation of the trap potentials discussed in Sec. \ref{sec:potentials}. 

\subsection{Step-by-step procedure for computing shuttling solutions}
\label{sec:stepbystep}
We describe how the framework can be employed for the particular use case of computing shuttling solutions for a given trap geometry.

\begin{figure}[!ht]
    \centering
    \includegraphics[width=0.9\columnwidth]{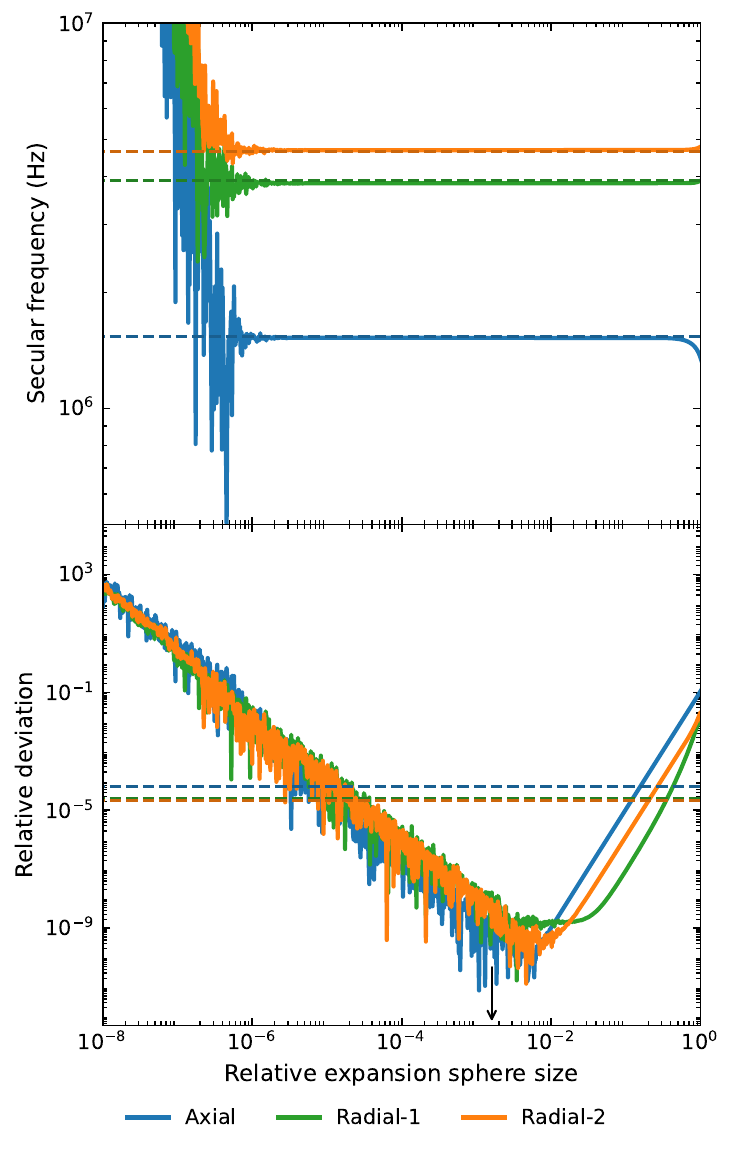}
    \caption{\justifying Secular frequencies versus expansion sphere size: The top panel shows the computed secular frequencies versus the relative expansion sphere size $\kappa/d$, with $d$ being the minimum ion-electrode distance used as characteristic length scale. The dashed lines indicate measurement results obtained with identical trapping parameters (see text). The bottom panel shows the deviations between measured and computed secular frequencies, referenced to the values obtained for the expansion radius marked by the arrow. The dashed lines indicate the relative accuracies of the spectroscopy measurements. The expansion sphere radii are indicated as normalized to a characteristic length of the trap geometry.
     }
    \label{fig:expansion_sphere_radius_scan_secular_frequency}
\end{figure}

\begin{enumerate}
    \item \label{step:0}
    The trap geometry is modeled  \textsc{Coreform Cubit} in the form of Python code, which also allows for the definition of parametric geometries. The geometry is imported to \textsc{Nullspace ES}, which is used for computing the distribution of surface charge distributions pertaining to +1V applied to each electrode (set), with all other electrodes at 0V. The resulting charge distributions are stored and managed by NullSpace. 
    \item \label{step:1}
    Defined  a shuttling path in the form of locations $\bs{r}_{w,t}$, as described in Sec. \ref{sec:sitcons}, where $w$ labels the potential wells and $t$ labels the steps. 
    \item \label{step:2}    
    Construct Fibonacci grids is constructed on expansion spheres of radius $\kappa$ around each location an $\bs{r}_{w,t}$ contained in the shuttling path. The coordinate axes should be aligned along the desired oscillation directions of the secular modes (see Fig. \ref{fig:sketchShuttlingPathPotentials} and Eq. \ref{eq:designPointsTransform}).    
    \item \label{step:3}    
    For each expansion sphere, each grid point and each relevant electrode (set), compute the values of unit potentials by querying Nullspace. For $K$ grid points, $W$ potential wells, $T$ shuttling steps and $N+1$ electrode (sets), in total $KWT(N+1)$ queries are required. 
    \item \label{step:4} 
    Carry out the procedure for coefficient determination described in Sec. \ref{sec:multipoleExpansion}. This results in sets of $(L+1)^2$ expansion coefficients $\{c_i\}$ for each $\bs{r}_{w,t}$ and each electrode (set).  
    \item  \label{step:5} 
    Specify the desired confinement along the shuttling path in the form of a set of target Hessians  $H^{(set,w,t)}$, by constructing diagonal matrices containing the desired secular frequencies $\omega_u$, using Eq. \ref{eq:secularFreqFromH} and subsequently rotate the Hessians onto the coordinate systems defined in step \ref{step:2}.
    \item  \label{step:6} 
    For each shuttling step, compute the quantities $e_{u,n}^{(w,t)}$ (Eq. \ref{eq:enFromCoeffs}), $h_{u,u',n}^{(w,t)}$ (Eq. \ref{eq:hnFromCoeffs}), $E_{\text{rf},u}^{(w,t)}$ (Eq. \ref{eq:ErfFromCoeffs}) and $H_{\text{rf},u,u'}^{(w,t)}$ (Eqs. \ref{eq:htildea},\ref{eq:htildeb}) from the expansion coefficients of the unit potentials.
    \item  \label{step:7} 
    Fill the penalty vectors and scale these with the global weights $\mc{W}_0^{(p)}$ (see Eq. \ref{eq:penaltyScalingFactors}).
    \item  \label{step:8} 
    Set up the linear system by filling the $A$ matrix according to Eq. \ref{eq:AmatrixFinal} and the vector $\bs{b}$ according to  Eq. \ref{eq:bvectorFinal}. Solve the linear system using algorithms suitable for symmetric band diagonal matrices such as sparse conjugate gradient solvers. The solution is set of optimum electrode voltages $V_{n,t}$.
    \item \label{step:9}
    Assess the quality of the solution by computing metrics such as the deviation of the well positions with respect to the desired values, the deviation of the secular frequencies, the orientation of the secular modes or the maximum voltages.
 \end{enumerate}
If the solution fails to meet predefined criteria such as minimum tolerable secular frequency deviation or voltage limits, several options exist: i) Increase the weight of the penalty corresponding to the criterion, resume at step \ref{step:7}, ii) change shuttling path, resume at step \ref{step:2}, or iii) change the trap geometry and restart at step \ref{step:0}.\\
As indicated in Fig. \ref{fig:toolchain_workflow}, our framework allows for efficient automatization of such optimization procedures. By extracting quantities from the solution which reflect the quality, a numerical feedback loop can be closed. Using parametric trap geometries in step \ref{step:0}, trap geometry parameters may be optimized such that specific shuttling operations are facilitated or even enabled by the geometry. We emphasize that the framework is optimized for efficiency along the complete pipeline, in order to offer the potential for parametric trap design studies.\\

\subsection{Numerical accuracy of the multipole expansion}
\label{sec:numericalaccuracy}
\begin{figure*}[!ht]
    \centering
    \includegraphics[width=1.0\textwidth]{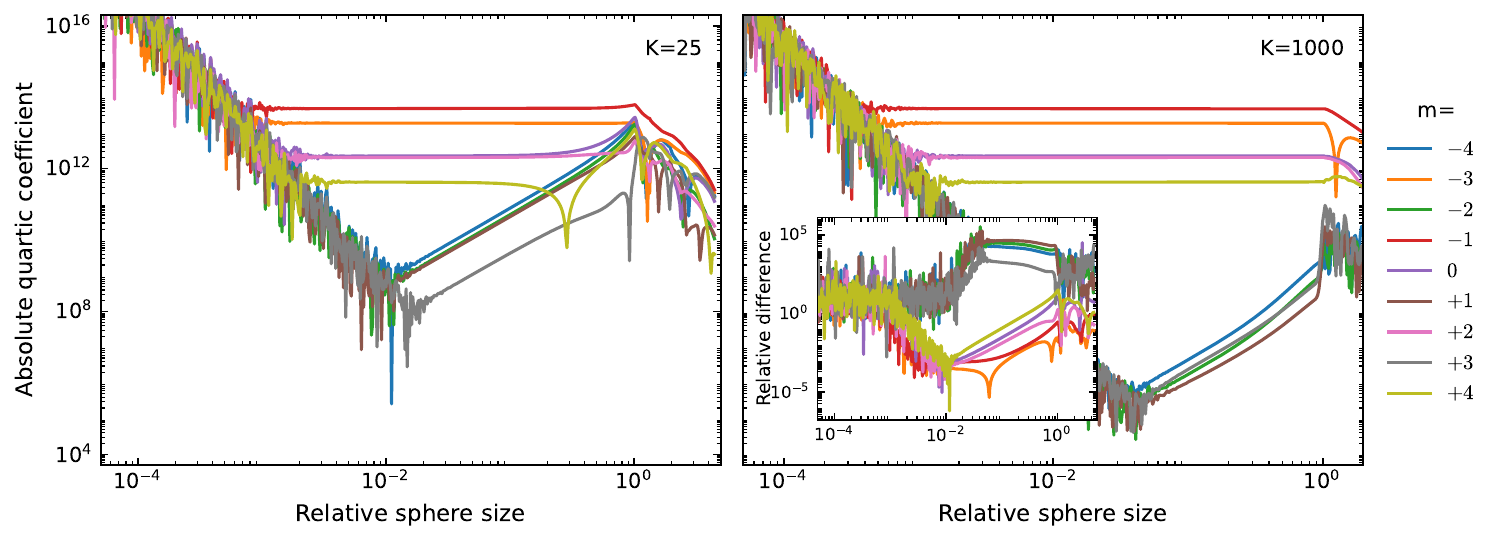}
    \caption{\justifying Anharmonic terms: The reconstructed solid harmonic coefficients for $l=4$ are shown for $L=4$ and  $K=25$ (left) and $K=1000$ (right) expansion points, versus the relative expansion sphere size $\kappa/d$ (see Fig. \ref{fig:expansion_sphere_radius_scan_secular_frequency}). The inset shows the moduli of the relative differences between the two cases.
    }
    \label{fig:expansion_sphere_radius_l_4_new}
\end{figure*}
We now assess the accuracy of the parts of the pipeline which are used to compute the expansion coefficients $c_i$ for representing the trap potentials by comparing computed and measured secular frequencies. The secular frequencies can be determined using laser spectroscopy, with relative accuracy and precision levels in the $10^{-5}$ range or even below, and therefore provide an excellent benchmark for validation of the numerical methods. We measure the axial and radial secular frequencies of a single $^{40}\text{Ca}^+$ in the linear, uniformly segmented trap employed e.g. in Refs. \cite{RUSTER2014,KAUFMANN2014,KAUFMANN2017,Hilder2022FTR}, using laser spectroscopy on the dipole-forbidden $S_{1/2}\leftrightarrow D_{5/2}$ transition near 729~nm. The ion is stored at the center of a trapping segment, with trapping parameters $\Omega=2\pi\times\SI{29.5}{\mega \hertz}$ and $V_{\text{dc}}=\si{- 6\volt}$ applied to the trapping segment. The secular frequencies are inferred from the difference frequency of motional sidebands with respect to carrier transition, such that the accuracy is limited by the ac Stark shift from the carrier and the precision is given by the width of the spectroscopic lines. The trap drive level $V_{\text{rf}}\approx\si{205 \volt}$ is inferred from the two measured radial secular frequencies.\\
Three parameters determine the accuracy of the computed secular frequencies:
\begin{itemize}
\item
    The tesselation of the trap geometry mesh used by the electrostatics solver in step \ref{step:0}. If the mesh is too coarse, the surface charge distributions may become the accuracy bottleneck. 
\item
    The radius $\kappa$ of the expansion sphere used in step \ref{step:2}. For too large radii, the truncation $L$ of the multipole expansion may become insufficient as higher order anharmonic terms become dominant. For too small values of $\kappa$, numerical errors are expected to occur.
\item 
    The number of grid points $K$ used in step \ref{step:2}.
\end{itemize}
\begin{figure}[!ht]
    \centering
    \includegraphics[width=\columnwidth,trim={0 0 0 0},clip]{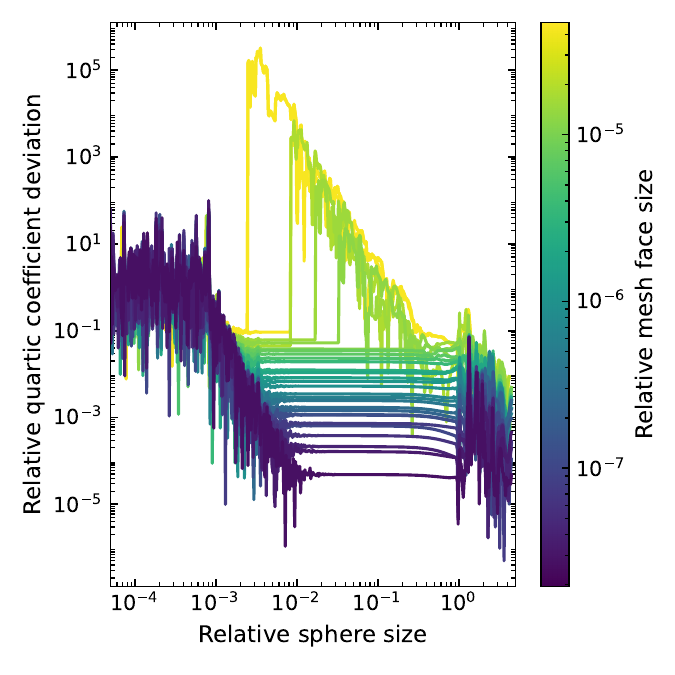}
    \caption{\justifying Impact of model tesselation on accuracy: Shown are the relative deviations of the expansion coefficient for $l=4,m=-1$ from Fig. \ref{fig:expansion_sphere_radius_l_4_new}, again versus the radius of the expansion sphere, for $K=25$ grid points and for different degrees of mesh fineness. The deviations are with respect to the coefficient value for the most fine mesh, and the fineness is expressed as the area of the smallest mesh face relative to the characteristic dimension $d^2 \approx \SI{5E4}{\micro\meter\squared}.$
    }
    \label{fig:expansion_sphere_model_mesh_scan_l4_m-1}
\end{figure}
For tesselation of the trap model, we use the \textsc{TetMesh} package of Coreform Cubit, generating a triangular mesh with decreasing structure size near edges and minimum face areas of \SI{27}{\micro\meter\squared}. For the multipole expansions, we employ $K=25$ grid points unless stated otherwise. For assessment of the numerical accuracy, we first study the $l=2$ multipole coefficients, which determine the secular frequencies via Eq. \ref{eq:hnFromCoeffs}, over a wide range of expansion sphere radii $\kappa$. The expansions are carried out at the center of the trapping segment where the ion is stored and for the trapping parameters as used for the measurements. The results are shown in Fig. \ref{fig:expansion_sphere_radius_scan_secular_frequency}. It can be seen that the computed secular frequencies for all three secular modes match the measurement values over wide range of $\kappa$ values, which are normalized to the minimal ion-electrode distance $d\approx \si{224 \mu \meter}$. For $\kappa/d \gg 10^{-6}$, we observe that the relative difference between computed and measured secular frequencies settles to $\lesssim 0.9\%$. For a more recent trap with similar geometry, fabricated using selective laser etching rather than the less accurate laser cutting and manual wafer stack alignment, we observe reduced mismatch levels of $\lesssim 0.3\%$ (data not shown). We conclude that the deviation levels do not reflect the actual accuracy of the computation, but rather are dominated by trap fabrication imperfections and stray fields.\\ 
We can nevertheless estimate the systematic errors of the computation pipeline: We shift all computed secular frequencies such that they exactly match the measured ones for a particular value of $\kappa/d=10^{-3}$ and reevaluate the deviations, it can be seen that the deviations remain consistently low (Fig. \ref{fig:expansion_sphere_radius_scan_secular_frequency}, lower panel). In the range $\kappa/d=10^{-5}\hdots 10^{-2}$, the frequency deviations for different values of $\kappa$ remain below the spectroscopic accuracy limit. The conclusion to be drawn from these results is that the computation is  numerically stable and that over a wide range of $\kappa$ values, the mismatch between measured and computed secular frequencies is \emph{not} determined by numerical errors. For $\kappa/d < 10^{-3}$, it can be observed that the solutions are affected by numerical artifacts from rounding, while for $\kappa/d > 10^{-2}$, systematic errors due to higher-order multipole terms become relevant.\\
Anharmonic terms of the trap potentials are important e.g. for separation and merging of ion chains \cite{Home2006,KAUFMANN2014} or for tailoring confinement properties of large Coulomb crystals \cite{Home2011}. Anharmonic potential coefficients are notoriously hard to extract from electrostatic simulation in a reliable fashion, and the results can hardly be benchmarked against measurement results. We systematically study the reliability of quartic ($l=4$) coefficients computed for the same electrode geometry as discussed above, again at the symmetry center of the trapping segment, the results are shown in Fig. \ref{fig:expansion_sphere_radius_l_4_new}. The coefficients pertaining to $m\in\{-4,-2,+1,+3\}$ are expected to vanish due to symmetry as they contain odd powers of the axial direction $x$, these indeed exhibit low values over a range of expansion radii around $\kappa/d\approx 10^{-2}$. The coefficients expected to have nonzero values exhibit plateaus around this value. Note that the quartic coefficients vary over several order of magnitude, and the spurious coefficient can assume the values of the smallest nonzero coefficient beyond $\kappa/d \approx 10^{-1}$.\\
As $K=25$ cannot produce a spherical $8$-design which would in principle be required for determination of the quartic coefficients, we also compute the coefficients for the case $K=1000$. We observe that the plateaus for the non-spurious coefficients extend to larger values of $\kappa/d$, because higher-order anharmonic terms - although not computed - are correctly captured. Moreover, the spurious coefficients are more reliably suppressed. Computing the relative deviations, we see that these remain way below 1\% around $\kappa/d=10^{-2}$. The conclusion to be drawn is that for most practical purposes, the choice $K=25$ and $\kappa/d=10^{-2}$ allows for extracting valid quartic coefficients. \\
We also study the impact of the fineness of the trap geometry mesh used by the electrostatic solver by computing the dominant anharmonic coefficient ($l=4, m=-1$, see Fig. \ref{fig:expansion_sphere_radius_l_4_new}) for different tesselations. The results are shown in Fig. \ref{fig:expansion_sphere_model_mesh_scan_l4_m-1}. It can be seen that for minimum face areas below about $10^{-5} d^2$, the relative deviations of the coefficient value with respect to the finest mesh remain below 10\%, for expansion sphere radii well within the accuracy plateau. For coarse tesselations however, the extracted anharmonic coefficients can deviate by several orders of magnitude from the true values, with strong dependence on the expansion sphere radius.\\ 
From the findings shown in the section, we conclude that for reasonable parametrizations, the accuracy of the expansion coefficients computed with our framework is not limited by the tesselation of the mesh used by the electrostatic solver. Furthermore, a small number of design grid points $K$ is sufficient for extracting coefficients for orders of up to $L=4$, which is one cornerstone of the numerical efficiency of our framework. The accuracy levels obtained are sufficiently good, meaning that the systematic errors become irrelevant for practical purposes, where the trap potentials are subject to stray fields or geometry imperfections. 
\begin{figure*}[t]
    \centering
    \includegraphics[width=0.9\textwidth]{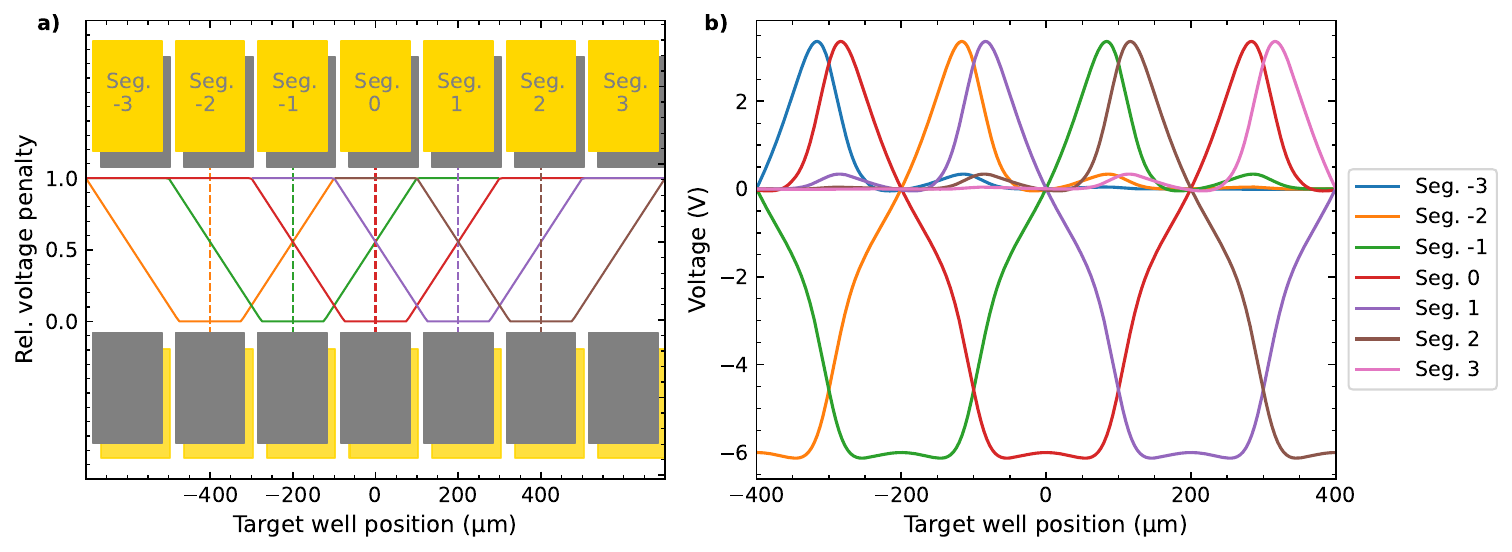}
    \caption{\justifying Linear multi-segment shuttling over five segments. \textbf{a)} Trap geometry and relative voltage penalty weight versus positions as determined by the activation functions Eq. \ref{eq:activationFunctions}. The shuttling path extends from the center of segment -2 to the center of segment +2. The dc electrodes (yellow) and rf electrodes (grey, tilted for visibility) are indicated. Dashed lines indicate the center positions of the segments used for computing the activation functions. \textbf{b)} Resulting control voltages versus position along the shuttling path for the five segments. 
    }
    \label{fig:multi_segment_shuttling}
\end{figure*}

\subsection{Performance}
\label{sec:performance}
Our framework allows for a wide range of possible applications based on its numerical performance, which becomes particularly relevant for trap geometries of increasing complexity. For the use case of computing shuttling solutions for a given trap geometry, the parametrization of the task in terms of the penalty vectors may have to be adapted until a satisfactory solution is found, which is facilitated by rapid computation of the solutions. More advanced use cases such as parametric trap design studies are only viable if the numerical framework features sufficient performance along the entire pipeline.\\
We exemplify the performance metrics for the task of computing a shuttling solution for shuttling a potential well around the corner of an X-type trap junction as illustrated in Fig. \ref{fig:sketchShuttlingPathPotentials} and presented in detail further below in Sec. \ref{sec:junction}. The trap model consists of 40 dc electrodes and one rf electrode, represented by 60748 triangular surface elements in total. The shuttling path consists of $T=300$ steps, for each of which multipole expansions for each electrode up to order $L=3$ based on $K=25$ grid points need to be computed. Nullspace is executed on a dedicated workstation (Lenovo P620 with 16 core AMD Threadripper 5955WX CPU featuring 16 cores and 384~GB RAM), and the remaining parts of the framework run on a standard desktop PC.\\
Solving the surface charge distribution (step \ref{step:1}) for the 40+1 on the workstation takes about 230~s. The multipole expansions for each electrode and each support point on the shuttling path (steps \ref{step:4}, \ref{step:5}) are computed within about 11~s. Note that the sets of points on which the potentials are computed and the resulting potential values (step \ref{step:3}) are communicated between the workstation and the client in the form of batches, in order to save network communication overhead, increase robustness against communication failures, and to harness the parallelization of NullSpace. If the expansion coefficients have already been computed, loading of these takes about 0.2~s. Parametrizing the shuttling task in terms specifying the fields and Hessians in terms of the multipole coefficients, the target Hessians and penalty vectors, including the segment activation functions, takes about 0.2~s (steps \ref{step:6},\ref{step:7},\ref{step:8}). Setting up the linear system $A,\bs{b}$ according to Eqs. \ref{eq:AmatrixFinal},\ref{eq:bvectorFinal}, conversion into sparse format and solution using a conjugate gradient solver finally takes about 0.9~s (step \ref{step:9}). Further significant performance gain can be expected by directly populating $A$ in sparse format.

\section{Application: Linear shuttling}
\label{sec:linearshuttling}
In this section, we demonstrate the method for computing shuttling solutions presented in Sec. \ref{sec:sitcons} for a simple scenario - shuttling an ion or ion chain along the trap axis of a linear uniformly segmented trap. We again use the trap model used in Refs. \cite{RUSTER2014},\cite{KAUFMANN2014},\cite{KAUFMANN2017},\cite{Hilder2022FTR}, and the task is to move a potential well by a distance of four trap segments of $\SI{200}{\micro\meter}$ width each, while keeping the variations of the axial and radial secular frequencies as low as possible. We follow the procedure described in Sec. \ref{sec:stepbystep} and use the same trapping parameters as in Sec. \ref{sec:numericalaccuracy}. The shuttling path is subdivided into $T=400$ steps, and the constant reference Hessian pertains to secular frequencies of about $2\pi\times\SI{1.57}{\mega\hertz}$ in the axial directions and  $2\pi\times\SI{3.86}{\mega\hertz}$, $2\pi\times\SI{4.73}{\mega\hertz}$ along the radial directions. The linear system $A,\bs{b}$ is set up as described in Sec. \ref{sec:linearSystem}, using the electrode activation functions Eq. \ref{eq:activationFunctions} within the voltage penalty. The $2800\times 2800$ linear system is solved using a sparse conjugate gradient solver to obtain the optimum voltage sample sets. The trap geometry as well as the resulting voltage solutions are shown in Fig. \ref{fig:multi_segment_shuttling}. Smooth, translationally symmetric voltage solutions are obtained, and the voltage levels fall within the $\pm$10~V range offered by typical arbitrary waveform generators.
\begin{figure*}[!htp]
    \centering
    \includegraphics[width=0.9\textwidth]{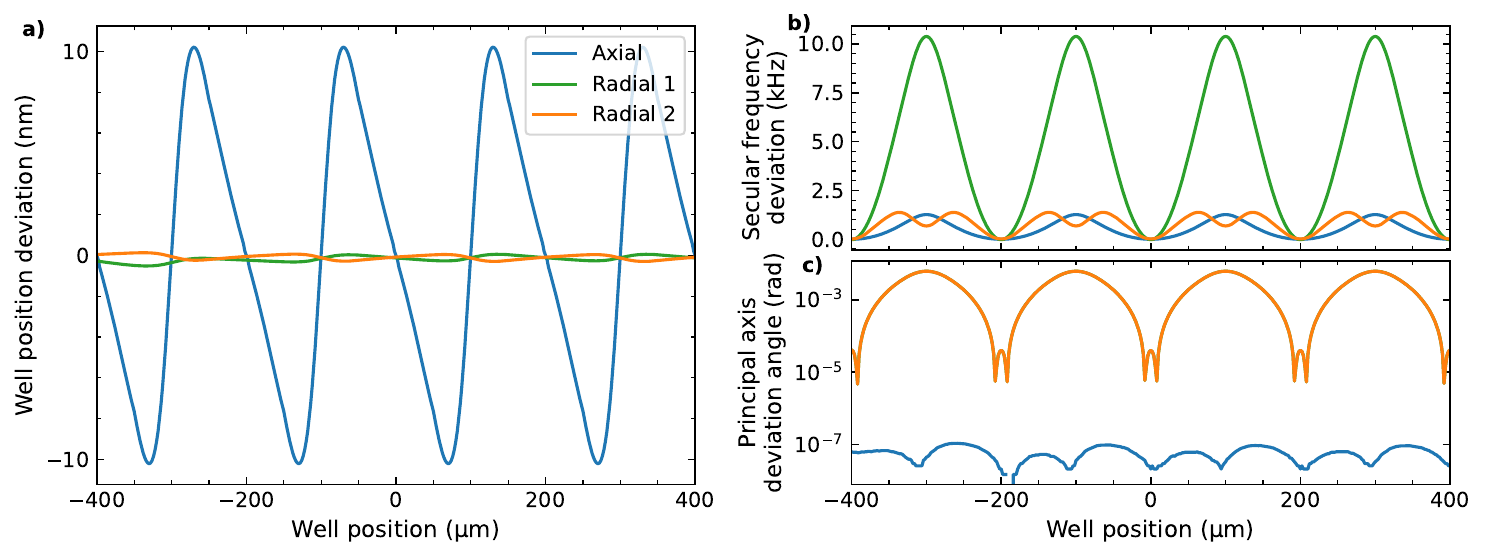}
    \captionsetup{justification=justified}
    \caption{\justifying Analysis of the shuttling solution for multi-segment transport. \textbf{a)} Deviations of the well positions from their preset values along the secular mode directions. \textbf{b)} Deviations of the secular frequencies, and  \textbf{c)} Angle deviations of the principal axes with respect to the preset directions.}
    \label{fig:multi_segment_shuttling_analysis}
\end{figure*}
In Fig. \ref{fig:multi_segment_shuttling_analysis}, we show the performance metrics of the shuttling solution. The deviations of the well positions along the shuttling path are computed according to
\begin{equation}
    \Delta r_{u,t} = \frac{Q\;E_{u,t}}{m\omega_{u}^2},
\end{equation}
where $Q\;E_{u,t}$ is the total force Eq. \ref{eq:totalforce} along direction $u$ and evaluated at well position $\bs{r}_t$, and $\omega_u$ is the preset secular frequency. The position deviations along the axial direction remain in the range of $\pm\SI{10}{\nano\meter}$, while the residual deviations along the radial directions range way below $\SI{1}{\nano\meter}$. The residual radial deviations are numerical artifacts, as these ideally vanish by symmetry. Computing the secular frequencies along the shuttling path from the Hessians, we observe that the secular frequencies exhibit deviations well below 1\%. Moreover, the orientation of the radial principal axes display deviations from the initial directions of up to one mrad, periodic with the segment structure. 

\section{Application: Shuttling through a trap junction}
\label{sec:junction}
\begin{figure*}[ht!p]
    \centering
    \includegraphics[width=\textwidth,trim={0 0 0 0},clip]{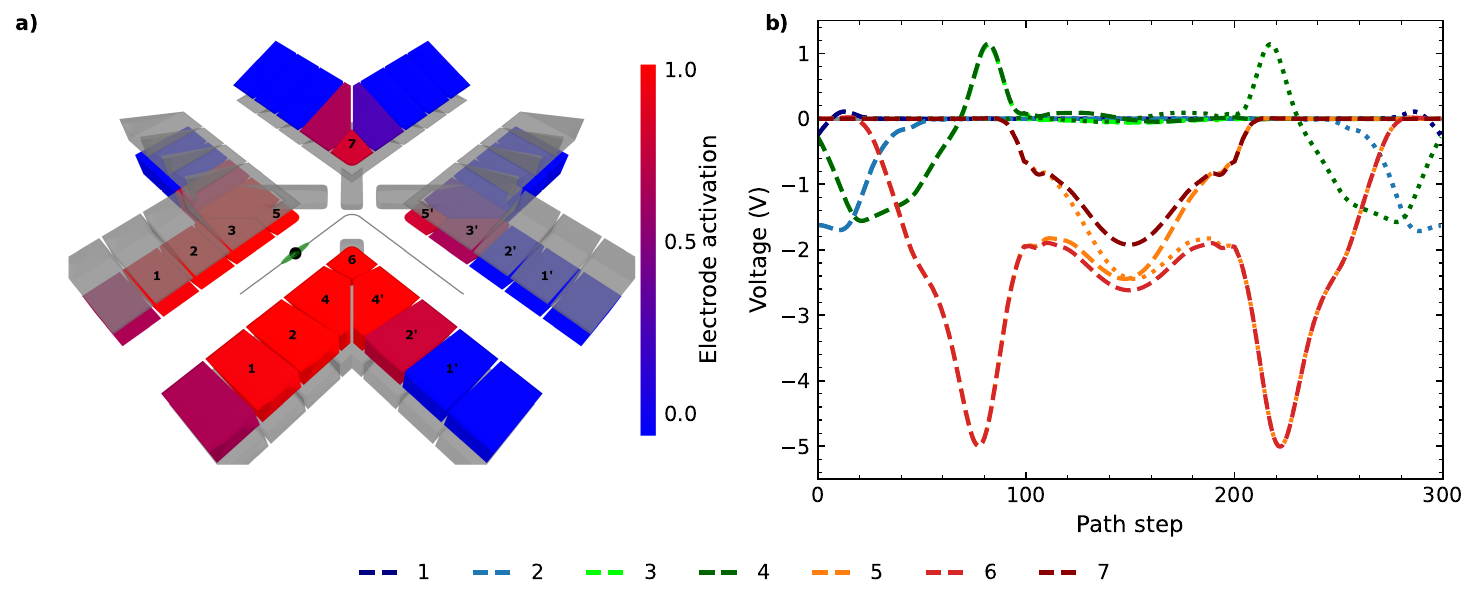}
    \caption{\justifying Shuttling around the corner of an X-type trap junction from \cite{Decaroli2021}. \textbf{a)} Trap geometry, the segments are color-coded according to activation functions, pertaining the indicated location at the shuttling path. The rf electrodes are shown in grey. \textbf{b)} Computed control voltages versus position along the shuttling path. In total 300 path steps are used, with a finer spacing for the 100 steps within junction region. The colors identify the electrodes via the numbers shown in a). The voltages for the electrodes in the lower-left (right, primed identifiers) arm are shown dashed (dotted). Voltages for remote electrodes assuming negligibly small levels are shown in grey.
    }
    \label{fig:junction_1}
\end{figure*}
Proposed architectures for scaling trapped-ion quantum processing units are based on traps featuring \emph{junctions} \cite{lekitsch2017,Malinowski2023}, i.e. trap regions such as depicted in Fig. \ref{fig:sketchShuttlingPathPotentials}, allowing for routing qubits into different trap regions. Junctions come in various types, with T-, X- or Y- type geometries. Shuttling through such trap element poses a challenge, as the ponderomotive confinement becomes weak within the junction and a ponderomotive force barrier needs to overcome by ions entering the junction region. The first successfully shuttling through a T-type being accomplished in 2006 \cite{Hensinger2006,Hucul2008}, significant progress has been achieved: Optimized electrode geometries have been developed both for stacked-wafer \cite{Wesenberg2009} and surface electrode traps \cite{Amini2010}. Shuttling through a stacked-wafer X-type junction was successfully demonstrated \cite{Blakestad2009} and subsequently improved to lead to low motional excitation \cite{Blakestad2011}. For surface electrode traps, both shuttling through X-type \cite{Wright2013} and Y-type \cite{Shu2014} junctions has been shown. Shuttling to any output port of an X-type junction in a surface electrode trap at motional excitations of $<0.1$ phonons on axial motional modes has been demonstrated recently \cite{Burton2023}, including mixed-species crystals. For upcoming intermediate-scale trapped-ion architectures, it can be anticipated that routing through junctions will be one key operation determining the platform capabilities, both in terms of timing resources and error budget. Efficient and accurate computation of shuttling solutions, particularly in view guiding trap designs, is therefore of vital importance. In this section, we show how shuttling solutions for a stacked-wafer X-type junction can be computed using our framework.\\
\begin{figure}[!htp]
    \centering
    \includegraphics[width=\columnwidth]{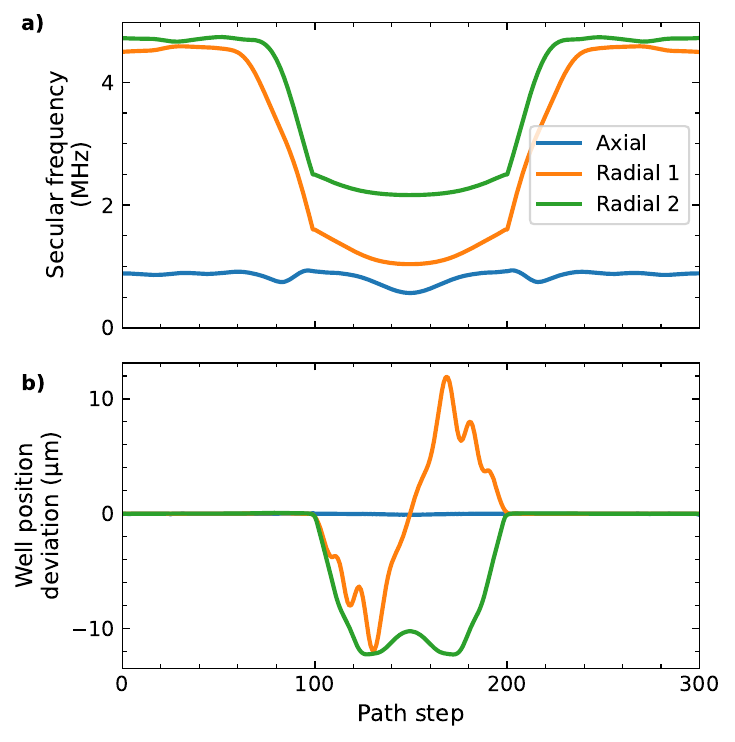}
    \captionsetup{justification=justified}
    \caption{\justifying Analysis of the shuttling solution for junction shuttling. \textbf{a)} shows the the absolute variation of the secular frequencies along the shuttling path. \textbf{b)} depicts the deviations of the well positions from their preset values along all three secular mode directions.}
    \label{fig:junction_2}
\end{figure}
We employ a trap geometry adapted from \cite{Decaroli2021}, limited to in total 36 dc control electrodes comprising the junction, see Fig. \ref{fig:junction_1}. Trap operation parameters and dimensions are the same as in \cite{Decaroli2021}. The junction features rf electrodes protruding into the junction in order to maintain stronger rf confinement. We define a shuttling path around a junction corner, comprised of $T=300$ points, with a $90^{\circ}$ bend of $\SI{50}{\micro\meter}$ radius at the junction center. The linear shuttling in both arms as well as the corner shuttling in the junction center consist of 100 steps each, the middle 100 steps being the corner shuttling. The target Hessian is defined to have axial direction at weakest confinement, remaining tangentially aligned to the shuttling path. All dc electrodes are independent (i.e. not grouped into sets of adjacent segment electrodes), and the electrode activation functions are parametrized such that the voltage penalty is effectively switched off once an electrode is fully activated. The shuttling solutions resulting from solving the band-diagonal 12000$\times$12000 linear system are shown in Fig. \ref{fig:junction_1}, it can be seen that all voltage levels remain well within reasonable bounds. From repeated computation of shuttling solutions, we found that only 16 electrodes contribute significantly to the shuttling solution. Furthermore, opposite electrodes in the linear regions can be grouped, such that only 12 independent control voltages remain, out of which five (primed electrodes in Fig. \ref{fig:junction_1}) exhibit mirror symmetry. Note while the edge electrodes (5,5',6,7 in in Fig. \ref{fig:junction_1}) are split in the original geometry from  \cite{Decaroli2021}, we found that this does not increase the quality of the control solutions, particularly in terms on radial confinement loss in the junction. Therefore, in our geometry, the respective electrodes are merged.\\
Fig. \ref{fig:junction_2} shows how the confinement properties vary along the shuttling path. As expected, substantial loss of radial confinement occurs close to the junction center, where the ponderomotive confinement is weakened. However, the variation of the axial secular frequency remains comparatively small, and the radial confinement remains stronger than the axial confinement along the entire path. While the mismatch of the axial placement of potential wells remains much smaller than $\SI{100}{\nano\meter}$, deviations along the vertical direction of up to about $\SI{10}{\micro\meter}$ are obtained. \\
While loss of radial confinement within a trap junction is generally unavoidable, several measures are available for mitigation, such as designing optimum electrode geometries \cite{Wesenberg2009,Amini2010,Zhang2022,taniguchi2025} and shuttling paths \cite{Wright2013}. It has to be taken into account that path deviations from the rf null help to salvage radial confinement, but may lead to increased heating from cross-coupled electric field noise \cite{Blakestad2009}. Our results show that using our framework, the performance of a junction geometry can be evaluated within a few minutes. Moreover, the results for a single parametrization of the shuttling task can be obtained within a few seconds once the charge distributions for the electrode geometry are solved (see Secs. \ref{sec:stepbystep},\ref{sec:performance}). The numerical efficiency of the toolchain may therefore help to conceive improved junction designs and corresponding shuttling protocols, allowing for controlled tradeoffs between design and control complexity and quality of the junction shuttling. 

\section{Postprocessing: From shuttling solutions to voltage waveforms}
\label{sec:postprocessing}
The shuttling solutions $V_{n,t}$ computed according to Sec. \ref{sec:sitcons} have to be post-processed in order to obtain actual \emph{voltage waveforms}, which can be transferred to a multichannel arbitrary waveform generator (AWG) \cite{kaushal2020}. The AWG, downstream electric circuitry as well as the trapped ions themselves are characterized by hardware parameters, i.e. sampling rate, bandwidth and inertia, which entails the need for matching the shuttling solutions to the hardware via postprocessing. The following discussion relies on the fact that all quantities defined with respect to sequence step $t$, such as the well positions $\bs{r}_{w,t}$, the expansion coefficients $c_i$ and the voltage samples $V_{n,t}$ can be turned into continuous functions via spline interpolations. Conversely, from an interpolation, a discrete sequence of values can be obtained.

\subsection{Mapping}
Having continuous shuttling solutions $V_n(\tau), \tau=0\hdots 1$ with $V_n(0)=V_{n,1},V_n(1)=V_{n,T}$ available from interpolation of a discrete shuttling solution, the next step is to determine a transfer function  $f \colon (0,1) \to (0,1)$ with $f(0)=0, f(1)=1$ such that mapped continuous shuttling solutions are computed from
\begin{equation}
    \tilde{V}_n(\tau)=V_n\left(f(\tau)\right).
\end{equation}
Various methods are available for design of such transfer functions, such such as invariant-based inverse engineering \cite{Torrontegui2011,Palmero2015}, where the final motional excitation is suppressed by harnessing the availability of closed-form solutions of the dragged parametric harmonic oscillator \cite{Lewis1969}.  Optimal control methods such as Krotov, GRAPE or CRAB approaches \cite{Koch2022} may also be employed, where optimal solutions are computed based on numerical solution of the classical \cite{Schulz2006} or quantum dynamics \cite{Singer2010}. These methods are beyond the scope of this work, we refer to \cite{Fuerst2014} for a comparative overview. \\
For shuttling operations in the adiabatic regime, it is generally sufficient to choose a simple sigmoid-type transfer function providing gradual acceleration and deceleration, such as e.g.
\begin{equation}
    f(\tau)=\sin^2\left(\frac{\pi \tau}{2}\right).
\end{equation}
After mapping, discrete waveforms are computed via resampling:
\begin{equation}
    V_{n,t}=\tilde{V}_{n}\left(\left(t-\tfrac{1}{2}\right)\Delta\tau\right) \quad t=1\hdots T,\;\Delta\tau=\frac{1}{T}.
    \label{eq:mappedResampling}
\end{equation}
Mind that we now use $V_{n,t}$ to denote the mapped, resampled voltages, and the step number $T$ may differ to the number of steps used for computing the original shuttling solution.

\subsection{Filter inversion}
\label{sec:filterInversion}
The voltage waveforms are supplied to the trap electrodes by an AWG, which will typically have output noise levels which would lead to non-tolerable heating of stored ions. Moreover, upon executing shuttling ramps, switching the output voltage levels leads to sharp spikes - termed glitches - which may also lead to large motional excitation. Passive or active filters in the signal lines are therefore required, see \cite{kaushal2020} for details. However, such filters will lead to delay and distortion of the signals, which degrades the control over the confinement properties. In particular, substantial buffer times may have to be added after shuttling operations in order to include the filter-induced ring-off behavior, which can have a detrimental impact on the overall system performance. While the voltage difference penalty Eq. \ref{eq:penalty4} used in the computation of the shuttling solution already reduces the required bandwidth, it remains beneficial to compute \emph{pre-ramps} from a mapped shuttling solution and the known properties of the filters, such that the output signal of the filter coincides as good as possible with the desired signal.\\
For the following discussion, we assume that nonlinearities of the AWG output stages remain negligible, and the effective linear filter is given by the output characteristics of the AWG and additional downstream circuitry. The filter is described by a discrete, finite convolution kernel and is therefore a finite impulse response (FIR) filter. Considering an input waveform $\{V_i\}$ (the electrode index is omitted) consisting of $T$ samples and kernel $\{k_j\}$ of length $K$, the output waveform $\{s'_j\}$ is given a convolution of the input waveform with the FIR kernel:
\begin{equation}
V'_i = \sum_{j=1}^{K} k_j V_{i-j+1}  \qquad i=1\hdots T.
\label{eq:firconvolution}
\end{equation}
It has to hold that $\sum_{j=1}^{K} k_j=1$, such that the convolution of a constant waveform leaves the output level unaffected. Our aim is now to cast Eq. \ref{eq:firconvolution} into the form of a matrix-vector product. We have to take into account possible ring-off behavior at the trailing edge, beyond $i=T$, and also additional samples before $i=1$ might be required for producing steep leading edges. We cast the input waveform into vector format and add some 'scratch space' by padding with $S$ copies of its first sample at the leading edge and $S$ copies of the last sample at the trailing edge:
\begin{equation}
\bs{V} = V_1 \bs{1}_S \; ; \; (V_1, \hdots, V_T) \; ; \; V_{T}\bs{1}_S
\end{equation}
where $';'$ denotes stacking of vectors, and $V_i\bs{1}_S$ denotes a vector consisting of $S$ times element $V_i$. The new waveform length is $M=T+2S$, and we assume that $S$ is chosen large enough such that $M>K$. The padding is justified for practical purposes, assuming the signal source remains idle sufficiently long before and after the waveform. This being fulfilled, we can safely describe the first $K$ samples of the filtered waveform by a set of kernel vectors $\bs{k}^{(m)}\; (m=K\hdots 1)$ which have a range of elements cumulated at the left (past kausal) end:
\begin{equation}
\bs{k}^{(m)}= \left(\sum_{j=K-m+1}^{K} k_j \;,\; k_{K-m+1}, \; \hdots \;, k_{1}\right).
\end{equation}
The length of $\bs{k}^{(m)}$ is $K-m+1$. This allows for writing the convolution Eq. \ref{eq:firconvolution} in matrix format:
\begin{equation}
\bs{V}'=\mc{K} \bs{V}
\end{equation}
with the $M \times M$ kernel matrix $\mc{K}$, specified in terms of its row vectors:
\begin{equation}
\mc{K}=\begin{pmatrix}
\bs{k}^{(K)} \;;\; \bs{0}_{M-1}\\ 
\bs{k}^{(K-1)} \;;\; \bs{0}_{M-2}\\ 
\vdots \\ 
\bs{k}^{(2)} \;;\; \bs{0}_{M-K+1} \\ 
\bs{k}^{(1)} \;;\; \bs{0}_{ M-K} \\ 
\bs{0}_{ 1}\;;\; \bs{k}^{(1)} \;;\; \bs{0}_{M-K-1} \\
\vdots \\ 
\bs{0}_{ M-K}  \;;\; \bs{k}^{(1)} 
\end{pmatrix}
\end{equation}
\begin{figure}[!htp]
    \centering
    \includegraphics[width=\columnwidth,trim={1.5cm 6.5cm 21.5cm 1cm},clip]{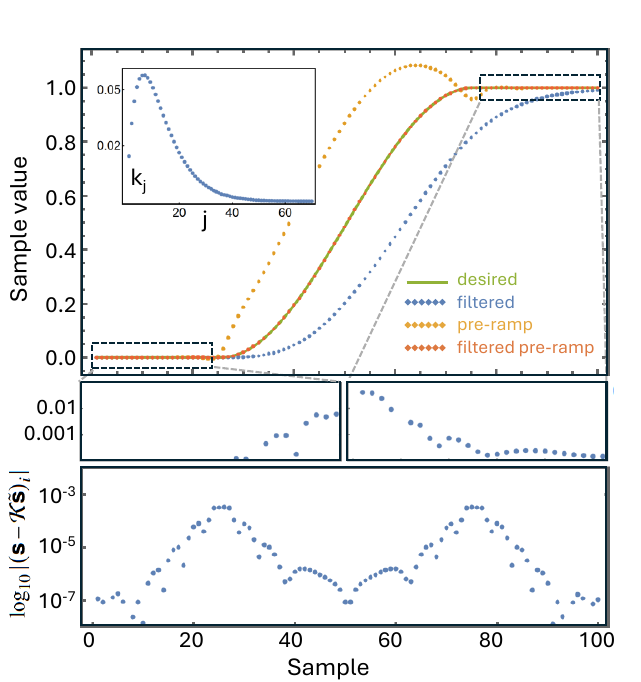}
    \captionsetup{justification=justified}
    \caption{\justifying Filter inversion for a $\sin^2$-type sigmoid waveform: The main panel shows a desired $T=50$ sample waveform (solid green) of $\sin^2$ type including padding with $S=25$ samples, and an toy-model $K=70$ kernel in the inset.  Also shown are the filtered (blue), pre-ramp (yellow) and filtered pre-ramp (orange) waveforms. The magnified regions show the modulus of the deviations of the pre-ramp and desired waveforms, indicating that most of the padded samples can be discarded. The lower panel shows the modulus of the filtered pre-ramp and desired waveforms, which remain well below the $10^{-3}$ level. A regularization parameter of $w=0.1$ was used. }
    \label{fig:filter_compensation}
\end{figure}
The inversion task now lies in finding an input ramp $\tilde{\bs{V}}$ such that a specified  ramp $\bs{V}$ is reproduced at the filter output, i.e. $\bs{V}=\mc{K}\tilde{\bs{V}}$. Attempting to solve directly by inversion of $\mc{K}$, i.e. $\tilde{\bs{V}}=\mc{K}^{-1}\bs{V}$ yields unsatisfactory results due to the fact that realistic low-pass filters lead to large condition numbers of $\mc{K}$, such that the inversion is not numerically stable and the solutions contain extreme voltage levels and bandwidth. We therefore need to include a suitable regularization. One approach is to enforce solutions minimizing the mean squared differences between consecutive voltage samples. The corresponding function we seek to minimize is therefore
\begin{equation}
\mc{F}(\tilde{\bs{V}})=\sum_{i=1}^{M} \left(V_i-\sum_{j=1}^{M}\mathcal{K}_{ij}\tilde{V}_j\right)^2 + w\sum_{i=2}^{M} (\tilde{V}_i-\tilde{V}_{i-1})^2 
\end{equation}
where $w>0$ is a hyperparameter weighing the difference penalty. We compute the derivative of $\mc{F}(\bs{\tilde{V}})$ with respect to sample $\tilde{V}_p$:
\begin{eqnarray}
\frac{d\mc{F}}{d\tilde{V}_p}&=&-2 \sum_{i=1}^{M}\mc{K}_{ip} \left(V_i-\sum_{j=1}^{M}\mathcal{K}_{ij}\tilde{V}_j\right) \nonumber \\
&-&2w (1-\delta_{p,1})(\tilde{V}_p-\tilde{V}_{p-1}) \nonumber \\
&-& 2w (1-\delta_{p,M})(\tilde{V}_{p+1}-\tilde{V}_p).
\end{eqnarray}
A stationary point of $\mc{F}$ is given if all derivatives vanish:
\begin{equation}
\frac{d\mc{F}}{d\tilde{\bs{V}}} = \bs{0}.
\label{eq:FIRinversionDerivative}
\end{equation}
Defining
\begin{equation}
    W_{ij}=\begin{cases}
        \delta_{i,j}-\delta_{i,j-1} & i=1 \\
        \delta_{i,j}-\delta_{i,j+1} & i=M \\
        2\delta_{i,j}-\delta_{i,j-1}-\delta_{i,j+1} & \text{otherwise,}
    \end{cases}
\end{equation}
Eq. \ref{eq:FIRinversionDerivative} can be stated as
\begin{equation}
\left(\mc{K}^T \mc{K}- \lambda W\right)\tilde{\bs{V}} = \mc{K}^T\bs{V}.
\end{equation}
from which we obtain the solution
\begin{equation}
\tilde{\bs{V}} = \left(\mc{K}^T \mc{K} - W\right)^{-1}\mc{K}^T\bs{V}.
\label{eq:preRampFinal}
\end{equation}
The kernel itself can be directly determined by recording the response to a unit voltage step on an oscilloscope, measuring at the output of a filter channel. This requires that no downstream passive components, such as additional capacitors near the trap, significantly affect the signals at the trap electrodes. The timing resolution is limited by the oscilloscope, therefore the resampling step of the mapped shuttling solution Eq. \ref{eq:mappedResampling} should match the resolution at which the step response is recorded. The kernel elements are then obtained by taking the differences of the recorded step response samples.\\
Results from the filter inversion method are shown in Fig. \ref{fig:filter_compensation}. A toy model kernel causing delay and distortion consisting of $K=70$ elements is applied to a $sin^2$ voltage waveform with $T=50$ samples. The chosen regularization weight $w=0.1$ leads a smooth resulting pre-ramp as computed from Eq. \ref{eq:preRampFinal}, and relative deviations between desired and filtered pre-ramp $\mc{K}\bs{\tilde{V}}$ remaining below the $10^{-3}$ level. Moreover, most of the used scratch space ($S=25$) can be safely discarded a-posteriori, as the relative sample levels also remain below $10^{-3}$. The final postprocessing step consists of resampling the pre-ramp to the timing resolution of the AWG.\\
The capabilities of the filter inversion method are ultimately limited by the slew-rate of the AWG output stages, i.e. the maximum voltage difference between consecutive samples \cite{kaushal2020}. Pre-ramps containing sample differences approaching the slew-rate limitation will lead to nonlinear signal distortions, such that the description in terms of a linear FIR filter breaks down.

\subsection{Mechanical simulation}
As a complementary tool, mechanical simulation of ion trajectories may be used for different purposes, e.g. to validate shuttling waveforms, to compute figure-of-merit metrics for closed-loop optimization of trap geometries or shuttling paths, or for identifying suitable total shuttling durations used in the mapping step for adiabatic shuttling. It is important to state that the scale separation between the ground-state wavefunction sizes for trapped ion (few nm to few tens of nm) and the size of control electrodes of (few tens to few hundreds of $\mu$m) renders the electrostatic and ponderomotive potentials to be almost perfectly harmonic across the extent of wavefunction. Therefore, for most scenarios, simulation of the classical point-mass dynamics is sufficient, even if ground state wavefunction and sub-quantum excitation energies are considered, see \cite{Fuerst2014}. \\
Within our framework, ion trajectories based on the multipole expansions of the trap potential along a shuttling path and a voltage waveform are computed using velocity Verlet integration \cite{swope1982}. The required prerequisites are as follows:
\begin{itemize}
    \item 
    Interpolations of all dc potential expansion coefficients up to order $l=2$ and rf expansion coefficients up to order $l=3$. Once the problem enforces motion only along the shuttling path, $l=1$ is for the dc potentials and $l=2$ for the rf potentials is sufficient.
    \item 
    Interpolation of all voltage waveforms, namely the filtered pre-ramps 
    \begin{equation}
        \mc{V}_n(\tau)=\mc{I}_{\mc{T}}\left[\mc{K}\bs{\tilde{V}}_n\right](\tau)
    \end{equation}
    where $\mc{I}_{\mc{T}}\left[\cdot\right]$ denotes interpolation of a sample waveform scaled to final time $\mathcal{T}$.
\end{itemize}
The velocity Verlet algorithm requires computing the acceleration $\bs{a}(\bs{r},t)$ acting on an ion of mass $m$ and charge $Q$ at any position $\bs{r}$ in the trap coordinate system, taken at any time $\tau=0\hdots\mc{T}$. For the general case where $\bs{r}$ can lie outside of a shuttling path, we first compute the shortest vector $\bs{r}_d$ connecting to a point $\bs{r}_p$ within the path, such that $\bs{r}=\bs{r}_p+\bs{r}_d$. The acceleration is then computed from evaluating the  electrostatic (Eq. \ref{eq:En}) and effective ponderomotive (Eq. \ref{eq:rfforce}) force fields as well as the ponderomotive (Eq. \ref{eq:Hrf}) and electrostatic (Eq. \ref{eq:Hn}) Hessians at $\bs{r}_p$:
\begin{equation}
    \frac{m}{Q}\bs{a}(\bs{r},t)=\bs{E}_{\text{rf}}-H_{\text{rf}}\bs{r}_d+\sum_n \mc{V}_n(t) \left(\bs{e}_n- h_n\bs{r}_d\right).
    \label{eq:simulation1}
\end{equation}
For multi-ion scenarios, the Coulomb repulsions ought to be taken into account as well. We note that the force in Eq. \ref{eq:simulation1} acting tangential to a shuttling path includes anharmonicities of the trap potential up to arbitrary order, limited by the accuracy of the interpolations. Perpendicular to the shuttling path, the electrostatic and ponderomotive forces are included only to first order. In cases where it remains unsure whether the pseudopotential approximation is valid, it is possible to analyze the validity via mechanical simulations resolving the trap drive. This requires replacing the acceleration due to the rf field by
\begin{equation}
    \frac{m}{Q}\bs{a}_{\text{rf}}(\bs{r},t)=V_{\text{rf}}\left(\bs{e}_{\text{rf}}-h_{\text{rf}}\bs{r}_d\right)\; \cos\left(\Omega t+\phi_{\text{rf}}\right).
\end{equation}
Simulation results such as final excitation energies significantly depending on the rf phase $\phi_{rf}$ indicate that the pseudopotential approximation may not be legitimate.
 
\section{Conclusion and Outlook}
In this work we have presented a numerical framework for the efficient and accurate computation of multipole expansions of electric potentials in segmented ion traps or electrode geometries serving other purposes. Combined electrostatic and ponderomotive forces as well as secular trap frequencies can be computed from the multipole coefficients. Based on this, we have shown how shuttling solutions and post-process voltage waveforms can be computed using an integrated pipeline of hardware-informed, efficient and versatile methods. The method for computing shuttling solutions was illustrated using the simple scenario of moving a single potential well along the axis of a linear segmented trap, and a more complex task consisting of moving a potential well around the corner of an X-type junction in a stacked-wafer trap.\\
The framework can be adapted and extended to address a variety of tasks in the context of shuttling operations. The numerical efficiency of the framework allows for parametric trap design studies, where trap geometry parameters can be systematically optimized. It is even possible to perform automated closed-loop design optimization by adapting electrode geometry parameters according to the quality of shuttling solutions for a given shuttling task. Possible extensions of the framework may include covering multi-species scenarios \cite{Home2013multi}, or architectures enabling shuttling operations via controllable rf drive levels \cite{valentini2025}. Upcoming novel fabrication methods such as 3D printing \cite{Xu2025} yield trap structures with increased accuracy of the electrode geometry. Furthermore, post-fabrication trap analysis via X-ray tomography \cite{Levine2023} may also allow for numerical trap models better matching the actual geometry. Such methods would allow for leveraging the accuracy of our framework to design fast, low-excitation shuttling operations, which may reduce the requirement of sympathetic cooling. In the context of scalability, our framework may be helpful for developing and optimizing sophisticated structures such as direct \cite{Akhtar2023} our auto-ponderomotive \cite{Seidling2024,Schmidtkaler2025} matter links between trap chips. \\
Finally, our framework can be adapted to target related quantum information processing platforms such a electrons in microwave traps \cite{Matthiesen2021} or atomic ions in Penning traps \cite{Jain2024}, and might also find applications beyond quantum information processing, such as for the development of compact mass spectrometers \cite{Snyder2016}, trapped-ion based optical frequency standards \cite{Keller2019}, spectroscopy of molecular ions \cite{Schiller2003} or chemistry in multipole ion traps \cite{Wester2009}.

\begin{acknowledgments}
We are grateful to Larissa Thorne for detailed review of the manuscript. We acknowledge financial support by the BMBF via VDI within the projects SYNQ, IQuAn and ATIQ,  by the European Union’s Horizon Europe research and innovation programme under grant agreement No 101114305 (“MILLENION-SGA1” EU Project), by the DFG under the SPP 2514 and by the Office of the Director of National Intelligence
(ODNI), Intelligence Advanced Research Projects Activity (IARPA), under the Entangled Logical Qubits program through Cooperative Agreement Number W911NF-23-2-0216. The views and conclusions contained in this document are those of the authors and should not be interpreted as representing the official policies, either expressed or implied, of IARPA, the Army Research Office, or the U.S. Government. The U.S. Government is authorized to reproduce and distribute reprints for Government purposes notwithstanding any copyright notation herein.
\end{acknowledgments}

\appendix
\section{Fibonacci design}
\label{sec:AppFibonacci}
For establishing a spherical design which approximates an exact spherical $t$-design as good as possible, we distribute $K$ points on the surface of a unit sphere in a Fibonacci grid:
\begin{eqnarray}
    z_k &=& 1-\frac{2k}{K-1} \qquad r_k=\sqrt{1-z_k^2} \quad \phi_k=k \pi (3-\sqrt{5}) \nonumber \\
    x_k &=& r_k \cos\phi_k \quad y_k=r_k\sin\phi_k
\end{eqnarray}
Here, $\pi (3-\sqrt{5})\approx 137.5^{\circ}$ is the golden angle. This leads to an evenly spaced and quasi-uniform distribution of the design points. \\

\section{Validity limits for nonzero micromotion}
\label{sec:rfdcValidity}
Our framework for computing shuttling solutions is based on simple superposition of electrostatic and ponderomotive forces, which relies is valid for trap operation parameters sufficiently deep in the stability regime of the trap, i.e. for all secular frequencies ranging way below the trap frequency. This is however only valid for potential well located at the rf null of the trap. The well-positioning penalty $\mathcal{F}^{(1)}$ Eq. \ref{eq:penalty1} only requires the total effective force to vanish at a well location, which can lead to well positions with nonzero (actual, not effective) rf field $\bs{\mathcal{E}}_{\text{rf}}$. Then, a number of further sufficient validity criteria eventually need to be considered:
\begin{itemize}
    \item 
    The excess micromotion must be small compared length scales on which the ponderomotive and electrostatic forces change. For a potential well with nonzero rf field  amplitude $\vert \bs{\mathcal{E}}_{\text{rf}}\vert$, the micromotion amplitude is given by
    \begin{equation}
        r_{\mu} = \frac{Q}{m\Omega^2} \vert \bs{\mathcal{E}}_{\text{rf}}\vert  =\frac{QV_{\text{rf}}}{m\Omega^2}\vert\bs{e}_{\text{rf}}\vert
    \end{equation}
    and should fall below the characteristic length scales on which the fields vary:
    \begin{eqnarray}
        r_{\mu} \ll \frac{\vert\bs{E}_{\text{dc}}\vert}{\left\vert\nabla\vert\bs{E}_{\text{dc}}\vert\right\vert} \label{eq:mmsmallnessdc}\\
        r_{\mu} \ll \frac{\vert\bs{\mathcal{E}}_{\text{rf}}\vert}{\left\vert\nabla\vert\bs{\mathcal{E}}_{\text{rf}}\vert\right\vert} \label{eq:mmsmallnessrf}
    \end{eqnarray}
    \item
    The net effect of the electrostatic forces over a micromotion cycle should remain negligible, which translates to
    \begin{equation}
        \frac{\vert\bs{E}_{\text{dc}}\vert}{\vert\bs{\mathcal{E}}_{\text{rf}}\vert} \ll 1
    \end{equation}
    \item
    For a shuttling process with dynamic potential well positions $\bs{r}_w(\tau)$, the micromotion envelope seen by each well should be slowly varying:
    \begin{equation}
        \left\vert \frac{d}{d\tau}\bs{r}_w(\tau)\right\vert \ll \omega\; r_{\mu}
    \end{equation}
    with $\omega$ being the minimum secular frequency.    
\end{itemize}
All criteria can be evaluated by computing fields, field gradients and secular frequencies based on the multipole expansions around the static well positions $\bs{r}_{w,t}$ or dynamic well positions $\bs{r}_w(\tau)$ as discussed in Sec. \ref{sec:mulitpoleExpansionEandH}. For Eq. \ref{eq:mmsmallnessdc}, we note that 
\begin{equation}
    \left\vert\nabla\vert\bs{E}_{\text{dc}}\vert\right\vert=\frac{\sqrt{\bs{E}_{\text{dc}}H_{\text{dc}}^2\bs{E}_{\text{dc}}}}{\vert\bs{E}_{\text{dc}}\vert}
\end{equation}
which can be evaluated using Eqs. \ref{eq:dcTotalFieldandH},\ref{eq:enFromCoeffs} and \ref{eq:hnFromCoeffs}. For condition Eq. \ref{eq:mmsmallnessrf}, one may use
\begin{equation}
    \left\vert\nabla\vert\bs{\mathcal{E}}_{\text{rf}}\vert\right\vert=V_{\text{rf}}\frac{\sqrt{\bs{e}_{\text{rf}}h_{\text{rf}}^2\bs{e}_{\text{rf}}}}{\vert\bs{e}_{\text{rf}}\vert}
\end{equation}
also in conjunction with Eqs. \ref{eq:enFromCoeffs} and \ref{eq:hnFromCoeffs}.\\
Note that the validity criteria listed here are sufficient conditions, and a shuttling solution partially violating the requirements may still work out in practice.\\
It is also important to note that an ion located a position where the ponderomotive force exhibits a gradient, it experiences additional heating due to electric field noise at the drive frequency offset by the secular frequency along direction $u$ \cite{Blakestad2009,Blakestad2011}:
\begin{equation}
    \dot{\bar{n}}_u=\frac{Q^2}{4m \hbar \omega_u} \left(\partial_u \Phi_{\text{rf}}\right)^2 \frac{S_{V_{\text{rf}}}(\Omega+\omega_u)}{V_{\text{rf}}^2},
\end{equation}
where $\bar{n}_u$ is the mean phonon number for the base secular mode along direction $u$, and $S_{V_{\text{rf}}}$ is the voltage noise spectral density of the trap drive voltage.

\section{Ponderomotive Hessians}
\label{sec:ApprfHessians}
In Eq. \ref{eq:Hrf}, the first part of the pseudopotential  Hessian $\tilde{h}_{\text{rf}}^{(a)}$ and the matrix elements for the second part of the pseudopotential Hessian $\tilde{h}_{\text{rf}}^{(b)}$ explicitly evaluate to
\begin{widetext}
\begin{eqnarray}
\tilde{h}_{\text{rf}}^{(a)}&=&\frac{15}{4\pi} \begin{pmatrix}
c_5^2+c_8^2+\left(c_9-\frac{1}{\sqrt{3}}c_7\right)^2 & c_6c_8+\tfrac{2}{\sqrt{3}}c_5c_7 & c_5c_6 -c_8\left(c_9+\tfrac{1}{\sqrt{3}} c_7\right) \\
c_6c_8+\tfrac{2}{\sqrt{3}}c_5c_7 & c_5^2+c_8^2+\left(c_9
+\frac{1}{\sqrt{3}}c_7\right)^2 & c_5c_8+c_6\left(c_9-\tfrac{1}{\sqrt{3}} c_7\right) \\
c_5c_6 -c_8\left(c_9+\tfrac{1}{\sqrt{3}} c_7\right) & c_5c_8+c_6\left(c_9-\tfrac{1}{\sqrt{3}} c_7\right) & c_6^2+c_8^2+\tfrac{4}{3} c_7^2
\end{pmatrix}
\label{eq:htildea} \\
\tilde{h}_{\text{rf},11}^{(b)} &=& -\frac{3}{2} \sqrt{7} \left(\sqrt{30} c_{10} c_{2}-\sqrt{2} c_{12} c_{2}-2
   \sqrt{3} c_{13} c_{3}-3 \sqrt{2} c_{14} c_{4}+2 \sqrt{5} c_{15}
   c_{3}+\sqrt{30} c_{16} c_{4}\right) \nonumber \\
\tilde{h}_{\text{rf},12}^{(b)} = \tilde{h}_{\text{rf},21}^{(b)} &=& \frac{3}{2} \left(\sqrt{14} c_{4}
   \left(c_{12}-\sqrt{15} c_{10}\right)+2 \sqrt{35} c_{11} c_{3}+\sqrt{14}
   c_{2} \left(c_{14}+\sqrt{15} c_{16}\right)\right)  \nonumber\\
\tilde{h}_{\text{rf},13}^{(b)} = \tilde{h}_{\text{rf},31}^{(b)} &=& 3\left(- \sqrt{35} c_{11}
   c_{2}- \sqrt{21} c_{13} c_{4}+2 \sqrt{14} c_{14} c_{3}+ \sqrt{35}
   c_{15} c_{4}\right)  \nonumber\\
\tilde{h}_{\text{rf},22}^{(b)} &=&   \frac{3}{2} \sqrt{7} \left(\sqrt{30} c_{10} c_{2}+3
   \sqrt{2} c_{12} c_{2}+2 \sqrt{3} c_{13} c_{3}+\sqrt{2} c_{14}
   c_{4}+2 \sqrt{5} c_{15} c_{3}+\sqrt{30} c_{16} c_{4}\right) \nonumber \\
\tilde{h}_{\text{rf},23}^{(b)} = \tilde{h}_{\text{rf},32}^{(b)} &=& -3
   \left(\sqrt{35} c_{11} c_{4}-2 \sqrt{14} c_{12} c_{3}+\sqrt{21}
   c_{13} c_{2}+\sqrt{35} c_{15} c_{2}\right) \nonumber \\
\tilde{h}_{\text{rf},33}^{(b)} &=&   -6 \left(\sqrt{14} c_{12} c_{2}+\sqrt{21} c_{13}
   c_{3}+\sqrt{14} c_{14} c_{4}\right)
\label{eq:htildeb}
\end{eqnarray}
\end{widetext}
It has to be noted that the computation of $\tilde{h}_{\text{rf}}^{(b)}$ requires multipole expansions of order at least up to $L=3$. However, for points which lie at the rf null of the trap, where $\bs{e}_{\text{rf}}=0$, this part of the rf Hessian vanishes and therefore does not need to be taken into account.

\bibliographystyle{apsrev4-1} 
\bibliography{main}

\end{document}